\documentclass[preprint,aps,prd,showpacs,superscriptaddress,nofootinbib,tightenlines]{revtex4}

\usepackage{amsfonts}
\usepackage{mathrsfs}
\usepackage{graphicx}
\usepackage{amsmath}
\usepackage{amssymb}
\usepackage{bm}
\usepackage{bbm}

\def\OMIT#1{}

\newcommand{\nn}{\nonumber}

\newcommand{\beq}{\begin{equation}}
\newcommand{\eeq}{\end{equation}}
\newcommand{\bqa}{\begin{eqnarray}}
\newcommand{\eqa}{\end{eqnarray}}

\begin{document}

\title{\mbox{}\\[10pt]
Color-singlet relativistic correction to inclusive $\bm{J/\psi}$
production associated with light hadrons at $\bm{B}$ factories}

\author{Yu Jia\footnote{E-mail: jiay@ihep.ac.cn}}
\affiliation{Institute of High Energy Physics, Chinese Academy of
Sciences, Beijing 100049, China\vspace{0.2cm}}
\affiliation{Theoretical Physics Center for Science Facilities, Chinese Academy of
Sciences, Beijing 100049, China\vspace{0.2cm}}
\affiliation{Kavli Institute for Theoretical Physics China,
Chinese Academy of Sciences, Beijing 100190, China}

\date{\today}
\begin{abstract}
We study the first-order relativistic correction to the associated
production of $J/\psi$ with light hadrons at $B$ factory experiments
at $\sqrt{s}=10.58$ GeV, in the context of NRQCD factorization.
We employ a strategy for NRQCD
expansion that slightly deviates from the orthodox doctrine, in that
the matching coefficients are not truly of ``short-distance" nature,
but explicitly depend upon physical kinematic variables rather than
partonic ones. Our matching method, with validity guaranteed
by the Gremm-Kapustin relation, is particularly suited for the inclusive
quarkonium production and decay processes with involved kinematics,
exemplified by the process $e^+e^-\to J/\psi+gg$ considered in this work.
Despite some intrinsic ambiguity affiliated with the order-$v^2$ NRQCD
matrix element, if we choose its value as what has been extracted from a recent
Cornell-potential-model-based analysis, including the relative order-$v^2$
effect is found to increase the lowest-order prediction for the integrated
$J/\psi$ cross section by about 30\%, and exert a modest impact on $J/\psi$ energy,
angular and polarization distributions except near the very
upper end of the $J/\psi$ energy.
The order-$v^2$ contribution to the energy spectrum becomes
logarithmically divergent at the maximum of
$J/\psi$ energy. A consistent analysis may require that these large end-point
logarithms be resummed to all orders in $\alpha_s$.
\end{abstract}


\maketitle

\section{Introduction}

Inclusive production of heavy quarkonium (especially $J/\psi$) in
various high-energy collider experiments has long been an
intriguing and interesting topic, to which a vast number of works
have been devoted~\cite{Brambilla:2004wf,Bodwin:2009}.
To date the mainstream of
theoretical investigations in this subject is based upon the
nonrelativistic QCD (NRQCD) factorization approach, a formalism
combining the effective-field-theory machinery together with the
hard-scattering factorization~\cite{Bodwin:1994jh}.
In the context of NRQCD factorization, the inclusive quarkonium
production rate can be expressed in a factorized form, that is, an
infinite sum of products of the perturbatively calculable partonic cross
sections and nonperturbative but universal NRQCD matrix elements. One
great virtue of this approach is that its predictions
can in principle be systematically improved.
This approach systemizes, and, extends, the conventional color-singlet model
(CSM). One striking, and, probably also disputable, ingredient of
NRQCD factorization is the so-called {\it color-octet mechanism}, that a heavy
quark-antiquark pair in a color-octet configuration created in a hard
process, is presumed to have non-negligible probability to
transition into a physical quarkonium state plus additional soft
light hadrons. Historically, the rapid popularity gained by this novel mechanism
is perhaps due to its economic explanation of the so-called
`$\psi^\prime$ surplus puzzle'~\cite{Braaten:1994vv}.

Although NRQCD factorization has enjoyed considerable successes in
many inclusive quarkonium decay and production processes,
it also faces some serious challenges. Most notably,
the recent Fermilab Tevatron measurement for $J/\psi$ polarization at
large $p_T$ seems to contradict with the benchmark predictions of the color-octet
mechanism, i.e. the increasingly transverse polarization of the
hadro-produced $J/\psi$ with increasing $p_t$~\cite{Cho:1994ih}.
Moreover, there is also problem for $J/\psi$ production in $e^+e^-$ collision experiments.
The color-octet mechanism also anticipates that an enhanced number of
$J/\psi$ populate near the maximum energy
region in $e^+e^-$ annihilation, but unfortunately, this quite distinct signature
has also not been confirmed by recent $B$ factory experiments.

These acute discrepancies have triggered a great wave of
theoretical efforts in recent years. Recent technical advancement makes it possible, for the first time,
to compute the rather involved next-to-leading QCD corrections to $J/\psi$ hadroproduction
in color-singlet channel~\cite{Campbell:2007ws,Artoisenet:2007xi,Gong:2008sn,Gong:2008hk,Gong:2008ft},
and its effects seem to be quite significant, i.e.,
to enhance the leading-order CSM contribution enormously.
This may indicate that, the phenomenological impetus to including color-octet contribution
seems not as indispensable as that in a decade ago, and the correct magnitudes of
color-octet matrix elements might be considerably smaller than the old numbers
extracted by implementing the LO CSM analysis only.

It is worth emphasizing that, the NRQCD factorization theorems for quarkonium production
are only at a conjectural level, which have never been proven rigourously to
hold to all orders in $\alpha_s$.
Notwithstanding the great utility of
the improvement on the short-distance coefficients, it is perhaps more urgent,
from the theoretical perspective, to reexamine every assumption underlying the
nonperturbative aspects of NRQCD factorization, especially for the color-octet mechanism.
As one of the important progresses along this line,
the validity of the factorization theorems
at two-loop level for gluon-to-quarkonium fragmentation
function has recently been established
after some suitable refinement of the original
color-octet NRQCD production operator~\cite{Nayak:2005rw}~\footnote{Now it becomes clear that
in some case NRQCD factorization certainly will fail.
For example, a novel phenomenon dubbed {\it color transfer}
mechanism~\cite{Nayak:2007mb}, was discovered in the production of
$J/\psi$ comoving with an additional heavy quark. In this case,
soft color exchange between the comoving quark and the constitutes of $J/\psi$
may invalidate the NRQCD factorization at two-loop level.}.
There is also suspect about the applicability of the
NRQCD velocity-scaling rule to charmonium, in particular it was suggested
that the spin-flip matrix element may play an important role
for the hardronization of color-octet $c\bar{c}$ pair~\cite{Liu:2006hc,Wu:2009ex}.
A more serious problem is that since each NRQCD matrix
element is a number instead of a distribution,  so NRQCD factorization
makes rather restrictive predictions to the various $J/\psi$ energy spectra.
It has long been suggested that in certain kinematic region
of quarkonium production, resummation of a class of
enhanced nonperturbative effects is crucial to make reliable prediction,
which effectively promotes the local NRQCD matrix element to a nonperturbative {\it shape function}~\cite{Beneke:1997qw,Mannel:1994xh}.
It is also worth noting that, there is also an ongoing endeavor to
circumvent the velocity expansion framework of NRQCD,
by introducing a more general set of fragmentation functions
in conventional perturbative QCD (pQCD) factorization base
to describe inclusive quarkonium production at large $p_t$~\cite{Qiu:2009}.

In recent years, $B$ factories also prove to be another active
field for the study of charmonium production. The
simplicity of the initial $e^+e^-$  state, together with the
enormous integrated luminosity, make the theoretical analysis of the
charmonium production process particularly clean and fertile.
Some recent measurements at $B$ factories have also posed challenges to
our understanding of charmonium production. One is the unexpectedly
large cross sections for several exclusive double charmonium production
processes in continuum $e^+e^-$ annihilation. For example, the cross
section for producing $J/\psi+\eta_c$ was first measured by the
\textsc{Belle} collaboration~\cite{Abe:2002rb}, which turns out to
be almost one order-of-magnitude larger than the leading order (LO)
NRQCD predictions~\cite{Braaten:2002fi}. After various theoretical
works from different angles, the consensus now is that
after including the large QCD perturbative correction~\cite{Zhang:2005ch,Gong:2007db},
in combination with relativistic corrections, this disquieting
discrepancy was claimed to be largely resolved
within the context of NRQCD factorization~\cite{He:2007te,Bodwin:2007ga}.

Another more perplexing observation arises from the inclusive production of
$J/\psi$. The production of $J/\psi$ in association with extra
charms, is found, quite counter-intuitively, to occur much more
copiously than that in association with a non-charm final state. The
fraction of number of events for $J/\psi$ plus charmed hadrons to
that of the inclusive $J/\psi$ events, conventionally denoted
$R_{c\bar{c}}$, was first measured by \textsc{Belle} collaboration
to be $0.59^{+0.15}_{-0.13}\pm 0.12$~\cite{Abe:2002rb},
later even shifted to $0.82\pm 0.15\pm 0.14$~\cite{Uglov:2004xa}.
This experimental values are in stark contrast to the leading-order (LO) NRQCD
predictions to this ratio, which is only about 0.1~\footnote{
Note that the LO NRQCD predictions to the $J/\psi$ production associated with charmed or
noncharmed hadrons are
identical to the CSM predictions~\cite{Keung:1980ev,Kuhn:1981jy,Driesen:1993us,Cho:1996cg,Yuan:1996ep,Baek:1998yf,Hagiwara:2004pf},
i.e. to proceed
through the parton processes $e^+e^-\to J/\psi+c\bar{c}$ and $e^+e^-\to J/\psi+gg$, respectively.}.
Other theoretical approaches, e.g., the estimate based on quark-hadron duality hypothesis~\cite{Berezhnoy:2003hz}
and the color-evaporation model~\cite{Kang:2004zj} also predict a quite small $R_{c\bar{c}}$.

Once upon a time, the total $J/\psi$ production rate measured at $B$ factories appeared
to be quite large, i.e., 2.5~pb measured by \textsc{Babar}~\cite{Aubert:2001pd}
and 1.5~pb~\cite{Abe:2001za} by \textsc{Belle}, which seems to request a sizable
color-octet contribution such as $e^+e^-\to c\bar{c}({}^1S_0^{(8)},{}^3P_J^{(8)})+g$.
The color-octet effect for $J/\psi$ production in $e^+e^-$ annihilation
was first investigated by Braaten and Chen~\cite{Braaten:1995ez} (see also \cite{Yuan:1997sn},
and for a very recent study of the NLO perturbative correction,
see \cite{Zhang:2009ym}.).
However, including this contribution will further dilute the ratio $R_{c\bar{c}}$.
Furthermore, an unusual signature of this mechanism is that an end-point peak is expected
in the $J/\psi$ energy spectrum. Unfortunately, there is no experimental evidences for
the existence of such a peak~\cite{Aubert:2001pd,Abe:2001za}.
To rescue the color-octet mechanism, later on the end-point Sudakov logarithms
have been identified and resummed, together with introduction of a
phenomenological shape function, one can show that the $J/\psi$ energy distribution
can be smeared out in accordance with the data~\cite{Fleming:2003gt}.
It is worth mentioning that absolute normalization of the color-octet contribution
is not affected much by including these refinements, and by that time
its contribution was assumed to predominate over the color-singlet contribution.

In the past couple of years, significant progresses toward resolving
these puzzles have been made from both experimental and theoretical angles.
From the experimental side, recently \textsc{Belle} collaboration was able to
precisely measure the cross sections for prompt $J/\psi$ production in association with
charmed and non-charmed states separately~\cite{Pakhlov:2009nj}:
\begin{subequations}
\bqa
\sigma[e^+e^-\to J/\psi+X_{c\bar{c}}] &=&  0.74\pm
0.08^{+0.09}_{-0.08}\;\,{\rm pb},
\\
\sigma[e^+e^-\to J/\psi+X_{light}] &=&  0.43\pm 0.09\pm 0.09\;\,{\rm
pb}.
\label{belle:latest:jpsi:plus:light}
\eqa
\end{subequations}
This new measurement has not subtracted
the feeddown contribution from $\psi(2S)$.

The most important recent theoretical progress in this subject is
perhaps the fulfillment of the NLO QCD corrections to both
channels. It turns out that the inclusion of the NLO QCD correction
significantly enhances the $J/\psi+c\bar{c}$ production rate~\cite{Zhang:2006ay,Gong:2009ng}.
Recently, the next-to-leading (NLO) perturbative correction to $e^+e^-\to J/\psi+gg$
has also been conducted, which enhances the LO cross section by only about 20\%~\cite{Ma:2008gq,Gong:2009kp}~\footnote{The end-point collinear logarithms
in the color-singlet channel has been resummed, but its impact on the $J/\psi$
spectrum seems rather insignificant~\cite{Lin:2004eu,Leibovich:2007vr}.}.
The significant enhancement to the former and the modest one to the latter
is of help for the predicted $R_{c\bar{c}}$ value to approach the measured one.
When the feeddown effects are included, the rough agreement
seems also to be achieved for both the associated $J/\psi$ production subprocesses,
so the alarming discrepancies
seem to be greatly alleviated~\footnote{However, although including
NLO QCD correction helps to get the right
answer for the inclusive production rate and energy spectrum of
$J/\psi$ in $e^+e^-\to J/\psi+X_{c\bar{c}}$,
it was noted that even ~\cite{Gong:2009kp}, it is still
difficult to reproduce the measured $J/\psi$ polarization
and angular distribution.}.

The above analysis tends to indicate that, CSM alone seems sufficient to
explain the data, and there seems no much room
left for the color-octet contribution.
As a result, the color-octet matrix element may be considerably smaller than
what was used to be assumed when fitted from the Tevatron data.
Nevertheless, it is still premature to assert that we already have
satisfactory understanding of inclusive $J/\psi$ production at $B$ factory,
because there is still one important component missing.
That is, in compliance with the NRQCD power counting,
one should also take the first-order relativistic correction
into account, since its effect is parametrically more important
than the color-octet contribution.
This is particularly relevant for $e^+e^-\to J/\psi+X_{light}$
since the size of NLO QCD correction to $e^+e^-\to J/\psi+gg$ is mild~\footnote{
The relativistic correction to $e^+e^-\to J/\psi+X_{c\bar{c}}$ has been
calculated and was reported to be surprisingly small~\cite{He:2007te}.}.
It is thus interesting to examine whether the relativistic correction
brings in sizable effects to this process or not.

The main purpose of this work is to answer this question,
that is, to calculate the first-order relativistic correction to
the inclusive $J/\psi$ production rate in
$e^+e^-\to J/\psi+X_{\rm light}$
in NRQCD factorization. Concretely, we will be considering the process
$e^+e^-\to J/\psi+gg$ at $O(\alpha_s^2)$.
It turns out that this correction is comparable in magnitude with
the NLO QCD correction, if not more important.

An experienced reader may agree that, calculations of QCD
perturbative corrections can be guided by some standard and
unambiguous procedure. By contrast, calculating relativistic
corrections, unexaggeratedly speaking, seems often plagued with
ambiguities and pitfalls~\footnote{See Ref.~\cite{Maksymyk:1997rq}
for an early discussion on one type of ambiguity affiliated with the
normalization of particle states in relativistic correction
calculations. See also \cite{Braaten:1998au} for a related discussion.}.
The problem gets particularly acute, when the
kinematics becomes involved, as in our case with three-body final states.
It often occurs that, different results have been reported by different authors in
calculating relativistic correction to the same
process~\footnote{For example, Ref.~\cite{Paranavitane:2000if}
claims a disagreement with an earlier publication~\cite{Jung:1993cd}
on the result of relativistic correction to photoproduction of
$J/\psi$. Likewise, a recent calculation of the relativistic corrections to
the fragmentation function for the $c$ quark to fragment into
$J/\psi$~\cite{Sang:2009} also disagrees with an earlier
work~\cite{Martynenko:2005sf}.}. To the best of our
knowledge, a simple and consistent recipe for calculating
relativistic corrections for a generic quarkonium production process
has not yet been explicitly given in literature.
One of the major motifs of this work is also to fill this gap.
We attempt to utilize a convenient yet slightly unconventional strategy
to deduce the relativistic corrections, applicable to any
inclusive quarkonium production (decay) process in color-singlet channel.
Our approach is slightly different from the
orthodox NRQCD doctrine, in that the matching coefficients are not
truly of ``short-distance" nature, for they explicitly depend upon
{\it physical} kinematic variables rather than the {\it partonic}
ones. However, we stress our method is still consistent, for its
validity resting upon a rigorous relation in NRQCD, the
Gremm-Kapustin relation~\cite{Gremm:1997dq}.

The rest of the paper is structured as follows. In
section~\ref{NRQCD:factor:long:dist:ME}, we state the NRQCD
expansion formula relevant to this work and discuss the
corresponding long-distance matrix elements of the color-singlet
production operators.
In section~\ref{NRQCD:matching:stragegy}, we outline our matching
scheme that can be applied to any color-singlet quarkonium
production process, and discuss its advantage over the more orthodox
doctrine.
In section~\ref{3:body:phase:space}, we give the differential expressions
for the three-body phase space needed
for the reaction $e^+e^- \to J/\psi+gg$.
In section~\ref{Description:matching:calculation}, we present a
detailed description on how to determine the short-distance
coefficients through relative order $v^2$ in $e^+e^-\to
J/\psi+X_{\rm light}$ and how the physical predictions for the
inclusive production rate of $J/\psi$ in $e^+e^-$ annihilation come
out.
In section~\ref{phenomena:Jpsi:distr:B:factory}, we apply our
formulas to investigate the phenomenological impact of the order-$v^2$
correction to the integrated production rate and the differential
energy spectrum for the unpolarized $J/\psi$, and the energy distribution
of angular and polarization parameters at $B$ factories.
Finally we summarize our results in section~\ref{summary:outlook}.
In Appendix~\ref{appendix:all:cross:section}, we collect the
analytic expressions for numerous types of differential cross
sections of $J/\psi$ in $e^+e^-$ annihilation, at the leading order
and next-to-leading order in $v^2$.
In Appendix~\ref{equiv:mine:matching:vs:orthdox}, we show that, it
is possible to reexpress our predictions for the integrated $J/\psi$
cross sections in a more orthodox form, that depending explicitly on
the charm quark mass rather than the $J/\psi$ mass.

\section{NRQCD factorization and long-distance matrix elements}
\label{NRQCD:factor:long:dist:ME}

According to the NRQCD factorization, the inclusive $J/\psi$
production rate in $e^+ e^-$ collision can be schematically written
as
\bqa
d \sigma[e^+e^-\to J/\psi+X] &=&  \sum_n d\hat \sigma [e^+e^-\to
c\bar{c}(n)+X] \langle {\cal O}^{J/\psi}_n \rangle.
\label{NRQCD:factorization:e+e-}
\eqa
The NRQCD expansion is organized by the velocity scaling of the
vacuum matrix element of NRQCD 4-fermion operators,
$\langle {\cal O}^{J/\psi}_n \rangle$, where $n$ denotes the
collective quantum numbers of the $c\bar c$ pair created
in the hard scattering.
The $d\hat \sigma_n$ are canonically referred to as the process-dependent
short-distance coefficients, which depend on all
{\it partonic} kinematic variables for a given production process,
but are insensitive to the long-distance aspects of
the quarkonium state $J/\psi$.

In this work we are only concerned with the color-singlet channel.
The readers who are also interested in the color-octet contributions to
this process can refer to Ref.~\cite{Braaten:1995ez,Yuan:1997sn,Fleming:2003gt}.
For our purpose, the relevant 4-fermion color-singlet production operators are
given by~\footnote{In this work, we find it convenient to choose a different
normalization for the $J/\psi$ production operators ${\mathcal O}_1^{J/\psi}$ and
${\mathcal P}_1^{J/\psi}$ other than the standard ones
introduced by Bodwin, Braaten and Lepage (BBL)~\cite{Bodwin:1994jh},
i.e., we define ${\mathcal O}_1^{J/\psi} \equiv {1\over 2J+1}{\mathcal
O}_{1\;{\rm BBL}}^{J/\psi}$ and ${\mathcal P}_1^{J/\psi} \equiv
{1\over (2J+1)\,m_c^2} {\mathcal P}_{1\;{\rm BBL}}^{J/\psi}$.  Note
the prefactor $1/m_c^2$ normalizes the operator ${\mathcal
P}_1^{J/\psi}$ such as to carry the same mass dimension as ${\mathcal
O}_1^{J/\psi}$.}
\begin{subequations}
\bqa
{\mathcal O}_1^{J/\psi}({}^3S_1) &=& {1\over 2J+1}\chi^\dagger
\bm{\sigma}\psi \sum_{X} |J/\psi+X\rangle \cdot \langle J/\psi+X|
\psi^\dagger \bm{\sigma}\chi,
\label{O1:jpsi}%
\\
{\mathcal P}_1^{J/\psi}({}^3S_1) &=& {1\over 2 m_c^2\,(2J+1)} \left[\,
\chi^\dagger \bm{\sigma}\psi \sum_{X} |J/\psi+X\rangle \cdot \langle
J/\psi+X| \psi^\dagger \bm{\sigma}
(-\tfrac{i}{2}\tensor{\mathbf{D}})^2 \chi +{\rm H.c.} \right],
\nn\\
\label{P1:jpsi}%
\eqa
\end{subequations}
where $\psi$ and $\chi$ are Pauli spinor fields for annihilating and a heavy quark,
and creating a heavy antiquark, respectively, $\sigma^i$ denotes
the Pauli matrix, and $\tensor{\mathbf{D}}$ is the spatial part of the
antisymmetrical covariant derivative: $\psi^\dagger
\tensor{\mathbf{D}}\chi\equiv \psi^\dagger {\mathbf{D}} \chi-
({\mathbf{D}} \psi)^\dagger \chi$, in a form to preserve Galilean
invariance. $J=1$ denotes the total spin of $J/\psi$, and $X$ denotes
additional light hadrons accompanied with $J/\psi$ with net energy
no larger than the ultraviolet cutoff of NRQCD. Note the sum for the intermediate states
is extended not only over all additional light flavor states $X$, but
over $2J+1$ polarizations of the $J/\psi$ as well.

The vacuum expectation values of these NRQCD production operators
are genuinely nonperturbative objects, whose exact values are even difficult
to ascertain from the powerful nonperturbative tools such as lattice QCD,
mainly owing to the obstacle in implementing those asymptotic states containing $X$.
Fortunately, in practice one can always invoke the so-called
{\it vacuum saturation approximation} (VSA) for the color-singlet channel,
which is accurate up to an error of relative order $v^4$,
to link these NRQCD operator matrix elements with the
more familiar Schr\"{o}dinger wave functions at the origin for the $J/\psi$:
\begin{subequations}
\bqa
\langle {\mathcal O}_1^{J/\psi} \rangle  &\approx&
\langle J/\psi|{\mathcal O}_1(^3S_1)_{\rm BBL} |J/\psi\rangle \approx  \left| \langle 0
|\chi^\dagger \bm{\sigma}\psi |J/\psi\rangle \right|^2
 = 2N_c \,\psi^2_{J/\psi}(0),
\label{O1:jpsi:vac:ME}%
\\
\langle {\mathcal P}_1^{J/\psi} \rangle &\approx& \langle J/\psi|{\mathcal
P}_1(^3S_1)_{\rm BBL} |J/\psi\rangle \approx    {\rm Re}\left[
\langle J/\psi|\psi^\dagger \bm{\sigma}\chi |0\rangle \cdot \langle
0 | \chi^\dagger \bm{\sigma} (-\tfrac{i}{2}\tensor{\mathbf{D}})^2
\psi|J/\psi\rangle \right]
\label{P1:jpsi:vac:ME}%
\\
&=& - 2N_c \,{\rm Re}\left[\psi^*_{J/\psi}(0)\overline{\nabla^2
\psi_{J/\psi}}(0)\right].
\nn
\eqa
\end{subequations}
Under this approximation, the vacuum matrix element for $J/\psi$ production
can be approximated by the square of vacuum-to-$J/\psi$ matrix element in NRQCD,
and further by the corresponding decay matrix element for the $J/\psi$ state.

The leading-order color-singlet $J/\psi$ production matrix element,
$\langle {\mathcal O}_1^{J/\psi}\rangle$, can be identified with the
familiar wave function at the origin for $J/\psi$, $\psi_{J/\psi}(0)$.
This quantity can be determined by several means, e.g., from lattice
simulation, or from phenomenological quark
potential models, or directly from the measured leptonic width of
$J/\psi$.

The determination of the relative order-$v^2$ production matrix element,
$\langle {\mathcal P}_1^{J/\psi}\rangle$, turns to be more problematic.
In Coulomb gauge, the gauge field piece in this matrix element
is suppressed relative to the ordinary derivative piece.
By VSA, it seems intuitive to interpret $\langle {\mathcal P}_1^{J/\psi}\rangle$
as product of $\psi_{J/\psi}(0)$ and the second derivative of the wave function at
the origin, $\nabla^2 \psi_{J/\psi}(0)$. Nevertheless, one should be
cautioned that such a naive interpretation is obscure. This is because
the bare NRQCD matrix element contains linear ultraviolet divergence,
hence it needs to be regularized and renormalized,
thus depending on the cutoff of the NRQCD lagrangian (An overbar put
above the wave function is to remind this).
There seems no direct way to directly infer this matrix element from
phenomenological potential model.
Nevertheless, it is the NRQCD effective theory
framework that endows this nonperturbative quantity
a meaningful definition.

For later use, it is convenient to introduce the dimensionless ratio of
the vacuum matrix elements of the following NRQCD operators:
\bqa
\langle v^2 \rangle_{J/\psi}  &=&  { \langle {\mathcal P}^{J/\psi}_1
\rangle \over \langle {\mathcal O}^{J/\psi}_1 \rangle } \approx
{\langle J/\psi(\lambda)| \psi^\dagger
(-\tfrac{i}{2}\tensor{\mathbf{D}})^{2}
\bm{\sigma}\cdot\bm{\epsilon}(\lambda)\chi|0\rangle \over m_c^2\,
\langle J/\psi(\lambda)| \psi^\dagger
\bm{\sigma}\cdot\bm{\epsilon}(\lambda)\chi|0\rangle}.
\label{v2:jpsi:definition}%
\eqa
This quantity, characterizing the typical size of
relativistic correction for $J/\psi$,
is supposedly around 0.3. Note that its value is independent of
the $J/\psi$ helicity $\lambda$ in above equation.

The vacuum-to-quarkonium relativistic correction matrix element
has been measured by lattice QCD, though the uncertainty is quite large.
There is an interesting relation,
first derived by Gremm and Kapustin (G-K)~\cite{Gremm:1997dq},
which derives from the equation of motion of NRQCD, and expresses
the relativistic correction NRQCD matrix element in terms of the LO
NRQCD matrix element, physical $J/\psi$ mass and the charm quark
mass:
\bqa
{M_{J/\psi}\over 2 m_c} &=&  1+ {1\over 2} \langle v^2
\rangle_{J/\psi}  + O(v^4),
\label{G-K:relation:first}%
\eqa
In NRQCD, the quark mass parameter is most naturally
identified with the quark pole mass. Unfortunately,
due to the intrinsic ambiguity of the charm quark pole mass,
this relation cannot be utilized to nail down the precise value
of $\langle v^2\rangle_{J/\psi}$.
It can not be precluded that, the actual value of this quantity might be
quite far from the naive expectation, 0.3. Since it is a subtracted quantity,
it will be perfectly consistent if $\langle v^2
\rangle_{J/\psi}$ turns to vanish or become even negative~\footnote{It is
worth mentioning that, recently there have been claims that this quantity can
be quite accurately extracted from the Cornell-potential-model-based
analysis,
$\langle v^2 \rangle_{J/\psi} = 0.225^{+0.106}_{-0.088}$~\cite{Bodwin:2007fz}.}.

During the era anteceding the inception of the NRQCD approach,
many authors preferred to using the
{\it binding energy}, i.e. $\epsilon\equiv M_{J/\psi}- 2 m_c$
to parameterize the contribution of relativistic corrections
(for example, see~\cite{Keung:1982jb,Jung:1993cd}).
With the aid of the G-K relation (\ref{G-K:relation:first}),
all those old results can be readily translated
into the modern form, with relativistic correction designated
by the NRQCD operator matrix elements.

\section{Perturbative matching strategy
\label{NRQCD:matching:stragegy}}

The central ingredient of the NRQCD factorization formula is
to deduce the NRQCD short-distance coefficients.
The procedure of determining these coefficients are usually referred to as
{\it matching}. The idea is rather straightforward, since these short-distance coefficients
are in principle insensitive to the long-distance nonperturbative physics,
therefore one may replace
$J/\psi$ by a free $c\bar{c}$ pair carrying the quantum number of ${}^3S_1^{(1)}$,
then both sides of Eq.~(\ref{NRQCD:factorization:e+e-}), including the NRQCD matrix elements in
the right side, can be accessed entirely in perturbation theory.
Matching both sides, one then be able to extract the desired short-distance coefficients
$d \hat{\sigma}_n$.

Let us take $e^+e^-\to J/\psi+X$ as an explicit example
to illustrate the problem faced for the matching calculation beyond the lowest order in $v^2$.
Schematically, one can express the corresponding perturbative matching formula for
(\ref{NRQCD:factorization:e+e-}) as
\bqa
\sum_X (2\pi)^4 \delta^4(K-P-k_X) \left| {\mathcal
M}\left[e^+e^- \to c\bar{c}(^3S_1^{(1)},P,\lambda)+X
\right]\right|^2 &=& \sum_n d \hat{\sigma}_n(P,\lambda) \langle
\mathcal{O}_n^{c\bar{c}} \rangle,
\label{NRQCD:matching:schematic}
\eqa
where the flux factor associated with the single-inclusive cross section has been
suppressed for simplicity. $K$ stands for the sum of momenta of the colliding electron and positron,
i.e., the 4-momentum of the virtual photon into
which the electron and the positron annihilate.
The sum in the left side is extended over the spins of all the additional partonic states $X$,
as well as over the phase space integration affiliated with $X$.

It is clear from (\ref{NRQCD:matching:schematic}) that, to identify the matching coefficients,
one needs to expand the inclusive production rate for perturbative $c\bar{c}(^3S_1^{(1)})$ pair
systematically in the small relative momentum between $c$ and $\bar{c}$, ${\mathbf q}$.
This procedure entails two essential ingredients, one is
to expand the matrix element squared in powers of ${\mathbf q}$, the other is
to expand the phase space integrals accordingly.
The former operation is more or less standard, but the
latter potentially cause some problems for the
orthodox matching method, as will be reviewed in section~\ref{matching:stragey:shape:function}.
The main trouble is that, in the standard matching calculation, it is $m_c$,
instead of the physical $J/\psi$ mass, $M_{J/\psi}$
that should enter the NRQCD short-distance coefficients.
The orthodox method then requires that the phase space integral be expanded around a fictitious
$c\bar{c}(^3S_1^{(1)})$ state of mass $2m_c$.
Technically, such an expansion of the phase space integral at differential level
is not easy to realize. More importantly, this operation leads to some inevitably
unsatisfactory feature in predicting the differential $J/\psi$ spectrum:
when approximating $J/\psi$ mass by $2m_c$,
the incorrect kinematics causes the spectrum somewhat distorted, which may become
particularly problematic near the phase space boundary~\footnote{If one is content to knowing only
the total cross section, this orthodox method should be straightforward and
does not cause any problem. For instance, for simpler reactions such as $gg\to \eta_c$,
exclusive double charmonium production $e^+e^-\to J/\psi+\eta_c$, or inclusive quarkonium decays $J/\psi\to ggg$, it is possible to first work out the phase space integration analytically,
then expanding the resulting integrated partonic production/decay
rates in powers of $\bf q$ about a fictitious charmonium of mass $2m_c$.
However, in the case of more involved kinematics, it is usually not feasible to acquire
the integrated rate in a closed form.}.

In section~\ref{matching:stragey:mine}, we will elaborate on the matching method
adopted in this work.
Motivated by the aforementioned shortcoming of the orthodox matching strategy,
we attempt to circumvent the most difficult part arising from
expanding the phase space integral. The key point is that we choose to
use physical kinematics instead of the partonic one, and the invariant mass of the $c\bar{c}$ pair
appearing in the matching calculation is taken as $M_{J/\psi}$.
As we shall see, this brings in great technical simplifications.
As a result, we can perform the matching at the level of the matrix element squared,
instead of at the level of the production rate. Although our matching method somewhat
deviates from the ordinary tenet in that the ``short-distance" coefficients now explicitly
depend on $M_{J/\psi}$, it is nevertheless still theoretically consistent,
thanks to the G-K relation (\ref{G-K:relation:first}).

\subsection{Orthodox matching strategy motivating the shape-function method
\label{matching:stragey:shape:function}}

In this subsection we review what the standard
NRQCD matching strategy would look like.
Historically, this method antecedes, and, motivates, the so-called
shape function method~\cite{Beneke:1997qw}.
The orthodox doctrine of NRQCD matching is common in any effective field theory,
in that the short-distance coefficients should depend only on the parton kinematics,
thus on the quark mass, and there is no way for quarkonium mass,
which inevitably entails the long-distance hadronization effect,
to enter into them.

Let $c$ and $\bar{c}$ that evolve to $J/\psi$ in (\ref{NRQCD:matching:schematic})
have momenta $p$ and $\bar{p}$.
Both $c$ and $\bar{c}$ are supposed to be on-shell, and
their momenta can be decomposed as
\begin{subequations}
\bqa
p&=& {1 \over 2} \widehat{P} + q_1,
\\
\bar{p} &=& {1 \over 2} \widehat{P} + q_2.
\eqa
\label{p-i:shape:func}%
\end{subequations}
Here the ``total" momentum $\widehat{P}$, which is deliberately chosen
to satisfy $\sqrt{\widehat{P}^2}=2 m_c$, should be distinguished from
the true total momentum of the pair, $P=p+\bar{p}$, with invariant
mass of $2 E_q$, where $E_q=\sqrt{m_c^2+\mathbf{q}^2}$ is the energy of the
$c$ or the $\bar{c}$ in the rest frame of the $c\bar{c}$ pair.
In the rest frame, these 4-momenta have
following explicit assignments:  $\widehat{P}^\mu=(2 m_c, \mathbf{0})$,
$q_1^\mu=(E_q-m_c,\mathbf{q})$, $q_2^\mu=(E_q-m_c,  -\mathbf{q})$,
respectively.
In the laboratory frame, it is understood that a suitable boost along the moving
direction of the $c\bar{c}$ pair is imposed on
all the 4-vectors.

The purpose of introducing $\widehat{P}$ is that one needs to expand the phase space integral
around a fictitious $c\bar{c}$ pair of invariant mass $2m_c$, which serves as the basis momentum.
Concretely, the constrained phase space measure for the partonic process
of (\ref{NRQCD:matching:schematic}) is
\bqa
d \Pi &=&  {d^3 \widehat{P} \over (2\pi)^{3}2 \widehat{P}^0}
\prod_i {d^{3}k_i \over (2\pi)^{3}2k_i^0}
(2\pi)^{4} \delta^{(4)}(K-\sum_i k_i - \widehat{P}- (q_1+q_2) ).
\label{phase:space:orthdox:matching}
\eqa
where $K$ is the momentum of the virtual photon,
and $k_i$ represents the additional partons in $X$. Note $q_i$ inside the $\delta$-function
are understood to be subject to an appropriate Lorentz boost from the rest frame
of $P$ to the laboratory frame.

The squared quark amplitude can then be folded with the phase space measure
(\ref{phase:space:orthdox:matching}) to obtain the partonic cross section.
The cross section needs to be expanded in the small momenta ${\mathbf q}_i$,
and powers of momentum can be identified with derivatives acting on the heavy quark
fields according to NRQCD factorization. Factors of relative momentum
$\mathbf{q}_1-\mathbf{q}_2$ (in the rest frame of the $c\bar{c}$ pair)
typically arise from expanding the amplitude, which can be identified
with the $\psi^\dagger (i\tensor{\mathbf{D}})\chi$. Furthermore,
in expanding the $\delta$-function
in phase space measure (\ref{phase:space:orthdox:matching}),
one typically encounters a different type of factor,
the center-of-mass (cms) type momentum $q^0_1+q^0_2$
(in the rest frame of the $c\bar{c}$ pair),
which can be identified with a total time derivative $i D_0(\psi^\dagger\chi)$.

As noted in Ref.~\cite{Beneke:1997qw} (see also \cite{Mannel:1994xh}),
these cms-derivative operators, though nominally of higher-order than the relative momentum
operators in NRQCD expansion, can be of dynamical significance near the kinematic endpoint of quarkonium
spectrum.
Upon expansion of the $\delta$-function in $q_1+q_2$, the resulting power series
in $v^2$ make increasingly singular contributions near the boundary of partonic phase space,
which signals that NRQCD expansion may break down near the endpoint region.
Fortunately, it has been shown that such enhanced kinematic contribution due to
these cms relativistic corrections can be resummed, whose effects are then encoded
in the universal nonperturbative {\it shape function}~\cite{Beneke:1997qw}.

The shape function is of greatest utility to improve the predictions for
inclusive quarkonium production in the
color-octet channel~\cite{Beneke:1997qw,Mannel:1994xh,Fleming:2003gt}.
Nevertheless, it can also play a nontrivial role even for the color-singlet channel,
which is relevant to our case.
It turns out that the resulting series from expanding the
$\delta$-function in (\ref{phase:space:orthdox:matching}) can be
exactly resummed without introducing any new nonperturbative parameter
other than the quarkonium mass.
Its sole effect is to account for the difference between quark and quarkonium mass,
consequently
shift the partonic boundary of phase space to the hadronic one.
The remarkable effect can be easily understood.
The cms-momentum factor $q_1^0+q_2^0$ equals $2E_q-2m_c$ in the rest frame of $P$.
When identified with the total time derivative
$i \partial_0 (\psi^\dagger\chi)$, this operator can convert to the binding energy
$\epsilon=M_{J/\psi}-2m_c$ when sandwiched between the vacuum and the physical $J/\psi$ states, thus helping to recover the hadronic kinematics. Not surprisingly,
the underlying reason is nothing but the G-K relation.

The role played by the color-singlet shape function seems to strongly suggest that,
the inconvenience brought in by the orthodox matching method, i.e.,
the procedure of expanding and reassembling of the phase-space $\delta$-function,
may be avoidable. As will be elaborated in more detail in next subsection,
one may just remain the physical kinematics intact in
(\ref{phase:space:orthdox:matching}) throughout the matching computation.
Lastly, we mention the fact that, somewhat ironically, there seems no any practical calculation of
the complete first-order relativistic correction for the
inclusive quarkonium production process that
is based on this orthodox matching tenet.

\subsection{Matching strategy adopted in this work
\label{matching:stragey:mine}}

In light of the complication inherent in the orthodox matching method,
in this section we are going to describe a different matching strategy,
which is suitable for {\it any} inclusive quarkonium production (decay)
process in the {\it color-singlet} channel. We will
take the reaction $e^- e^+ \to J/\psi+gg$ as an explicit example to illustrate
our method.

When assigning the momenta of $c$ and $\bar{c}$
in perturbative matching, the separation between ``total" and ``relative"
momenta is just a matter of taste, by no means unique.
Here we will choose a different one from that in
section \ref{matching:stragey:shape:function}.
The momenta of the on-shell $c$ and
$\bar c$ can be decomposed in the following form:
\begin{subequations}
\bqa
p &=& {1 \over 2}P+q,%
\label{c:momentum:def}
\\
\bar{p} &=& {1 \over 2}P-q.%
\label{cbar:momentum:def}
\eqa
\label{p-pbar-mom}%
\end{subequations}
We stress here $P$ represents the true total momentum of the pair, $P=p+\bar{p}$, with invariant
mass of $2 E_q$, and now $P$ and $q$ are chosen to be orthogonal: $P \cdot q=0$,
in contrast to the choice made in (\ref{p-i:shape:func}).
In the rest frame of the  $c\bar{c}$ pair, the explicit components of the
momenta are $P^\mu=(2E_q,\mathbf{0})$,
$q^\mu=(0,\mathbf{q})$, $p^\mu=(E_q,\mathbf{q})$, and
$\bar{p}^\mu=(E_q,-\mathbf{q})$, respectively.
In the laboratory frame, one has $P^\mu=(P^0, {\mathbf P})=(\sqrt{{\mathbf
P}^2+4E_q^2},{\mathbf P})$ and appropriate Lorentz boost is understood
to be imposed on any other 4-vector.
It is worth mentioning that, it is this form of momentum assignment,
rather than (\ref{p-i:shape:func}), that
has been practically used in most calculations of relativistic corrections~\cite{Bodwin:2003wh,He:2007te,Sang:2009,Fan:2009zq,Bodwin:2002hg,He:2009uf}.

The greatest advantage of this kind of momentum decomposition is that,
there is no need to expand the total momentum $P$ of the $c\bar{c}$ pair around a
basis momentum with invariant mass of $2m_c$,
and we will leave phase space measure intact by assuming the $c\bar{c}$
pair with an invariant mass $2E_q$.
We argue by this way the relativistic effects in phase space measure
are automatically incorporated, at least to relative order $v^2$.
We will come back to the connection between the factor $E_q$
and $M_{J/\psi}$ in nonrelativistic expansion later in
this subsection.

Even though we are coping with {\it inclusive} $J/\psi$ production process,
but insofar as the color-singlet channel is concerned,
it is not necessarily be committed to the cross section level at the very beginning.
In fact, since we no longer need worry about the complication from the phase space
integral, it seems legitimate to invoke the NRQCD factorization at the amplitude level,
To this end, we need only retain those operator matrix elements that connect the
vacuum to the color-singlet $J/\psi$ state. To the order of desired
accuracy, the amplitude can be written as
\bqa
\mathcal{M}[J/\psi(\lambda)+gg] &=& \sqrt{2 M_{J/\psi}} \left[
F_0(\lambda) \langle J/\psi| \psi^\dagger
\bm{\sigma}\cdot\bm{\epsilon}\chi|0\rangle \right.
\nn \\
& + & \left. {F_2(\lambda)\over m_c^2}\langle J/\psi| \psi^\dagger
\bm{\sigma}\cdot\bm{\epsilon}
(-\tfrac{i}{2}\tensor{\mathbf{D}})^{2}\chi|0\rangle + \cdots
\right],
\label{nrqcd-exp:CS:ampl}%
\eqa
where $F_i(\lambda)$ are the corresponding short-distance
coefficients, which are Lorentz scalars formed by various kinematic
invariants in the reaction.
In particular, they also depend explicitly on the helicity of
$J/\psi$, $\lambda$. For the Lorentz-invariant amplitude in the
left-hand side of Eq.~(\ref{nrqcd-exp:CS:ampl}), ${\mathcal
M}[J/\psi(\lambda)+gg]$, it is most natural to assume
relativistic normalization for each particle state, since the
squared amplitude needs to be folded with the relativistic phase
space integral to obtain the physical cross section. However, in the
right-hand side of Eq.~(\ref{nrqcd-exp:CS:ampl}), the $J/\psi$ state
appearing in the NRQCD matrix elements conventionally assumes the
nonrelativistic normalization. To compensate this difference, one
must insert a factor $\sqrt{2 M_{J/\psi}}$ in the right side of
(\ref{nrqcd-exp:CS:ampl}).

Squaring both sides of (\ref{nrqcd-exp:CS:ampl}), summing over the
final-state spin/colors as well as averaging upon the initial-state
spins, the matrix element squared reads
\bqa
\overline{\sum} \, \left|{\mathcal M}\big[J/\psi(\lambda)+gg\big]
\right|^2 &=& 2M_{J/\psi} \langle {\mathcal O}^{J/\psi}_1 \rangle
\overline{\sum} \, \left\{ \left| F_0 \right|^2 + 2 \,{\rm Re}\left[
F_0 F_2^* \right] \langle v^2 \rangle_{J/\psi}+\cdots \right\},
\label{ampl:squared:jpsi:gg}%
\eqa
where the VSA has been invoked and
$\langle {\mathcal O}^{J/\psi}_1 \rangle$ has been given in
(\ref{O1:jpsi:vac:ME}), and $\langle v^2 \rangle_{J/\psi}$ defined
in (\ref{v2:jpsi:definition}). The symbol $\overline{\sum}$
indicates the suitable color-spin summation/average.

To determine the coefficients $|F_0|^2$ and $F_0 F_2^*+{\rm H.c.}$, we
follow the moral that these short-distance coefficients are insensitive to the
long-distance confinement effect, so one can replace the physical
$J/\psi$ state by a free $c\bar{c}$ pair of quantum number
${}^3S_1^{(1)}$, by which the NRQCD operator matrix elements can
be perturbatively calculated. The short-distance coefficients $F_i(\lambda)$ can
then be read off by comparing the full QCD amplitude for producing $c\bar{c}({}^3S_1^{(1)})$
and the corresponding NRQCD factorization formula.
In our case, the amplitude for producing a
$c\bar{c}({}^3S_1^{(1)})$ pair associated with two gluons is~\footnote{Throughout this work,
the bold-faced symbols, such as $\mathbf q$, if not otherwise
stated, are exclusively referring to the spatial vectors defined in
the rest frame of the $c\bar{c}(P)$ pair, whereas the italic
symbols, such as $q$, are reserved for the covariant 4-vectors,
often presumed in the laboratory frame.}
\bqa
\mathcal{M}[c\bar{c}(^3S_1^{(1)},\lambda)+gg ] &=& F_0(\lambda)
\langle c\bar{c}(^3S_1)| \psi^\dagger
\bm{\sigma}\cdot\bm{\epsilon}\chi|0\rangle + {F_2(\lambda)\over
m_c^2} \langle c\bar{c}(^3S_1)| \psi^\dagger
\bm{\sigma}\cdot\bm{\epsilon}
(-\tfrac{i}{2}\tensor{\mathbf{D}})^{2}\chi|0\rangle
\nn \\
&=& \sqrt{2 N_c}\,2E_q \left[ F_0(\lambda)+  F_2(\lambda) \,
{{\mathbf q}^2\over m_c^2} \right].
\label{ampl:ccbar}%
\eqa
In Eq.~(\ref{ampl:ccbar}), we use relativistic normalization for the $c$
and $\bar c$ states in the computation of the full QCD amplitude
and in the computation of the NRQCD matrix elements.
Consequently, a factor $2E_q$ appears in the
second expression of Eq.~(\ref{ampl:ccbar}). An additional factor
$\sqrt{2N_c}$ arises from the spin and color factors of the NRQCD
matrix elements. From (\ref{ampl:ccbar}), it is straightforward to
extract the short-distance coefficients $F_i(\lambda)$:
\begin{subequations}
\bqa
F_{0}(\lambda) &=&  \left.
{ \mathcal{M}\big[c\bar{c}(^3S_1^{(1)},\lambda)+gg\big]\over \sqrt{2 N_c}
2E_q}\right|_{{\bf q}^2=0},
\label{F0:definition}
\\
F_2(\lambda) &=&  \left. {m_c^2 \over \mathbf{q}^2} \left(
{\mathcal{M}\big[c\bar{c}(^3S_1^{(1)},\lambda)+gg\big]\over \sqrt{2 N_c}
2E_q} - F_{0}(\lambda)\right) \right|_{{\bf q}^2=0}.
\label{F2:definition}
\eqa
\label{F0:and:F2:definition}
\end{subequations}
The LO coefficient $F_0$ can be obtained
by putting ${\mathbf q}\to 0$ in the amplitude and
equating $E_q$ and $m_c$.
To deduce the coefficient $F_2$, it is understood that one
has to first expand the amplitude to the first order in ${\mathbf q}^2$
prior to taking the ${\mathbf q}\to 0$ limit. Consequently,
it is necessary to distinguish between $E_q$ and $m_c$.

Although the expression of $F_0$ can be unequivocally determined, it
is not without ambiguity to deduce the coefficient $F_2$. This is
because, determination of this relativistic correction coefficient
crucially hinges on which quantity is chosen to be expanded around
in powers of ${\mathbf q}^2$ in the quark amplitude.

Obviously, those terms that contain explicit ${\mathbf q}^2$ factor
should contribute to $F_2$. Besides these terms,
in the matching procedure adopted by most authors,
one usually often includes
relativistic effects implicit in all the expressions that
contain the factor $E_q$, where $E_q$ is always expanded around $m_c$
in power series of $\mathbf{q}^2$, i.e. $E_q = m_c + {{\mathbf
q}^2\over 2 m_c}+{\mathcal O}({\mathbf q}^4)$. By collecting all the
sources proportional to ${\mathbf q}^2$, one is then able to deduce
the coefficient $F_2$ according to (\ref{F2:definition}).

In this work, we find it much more advantageous to take a somewhat
different route. Aside from retaining those relativistic
correction contributions that contain ${\mathbf q}^2$ explicitly,
we choose to expand every occurrence of $m_c$ in the
amplitude in term of ${\mathbf q^2/E_q^2}$, while keeping $E_q$
intact:
\beq
m_c = E_q - {{\mathbf q}^2\over 2 E_q}+{\mathcal O}({\mathbf q}^4).
\label{expand:m:in:Eq}%
\eeq
Somewhat nonstandard as it seems,
but as we will see shortly,
by choosing this way of expansion,
we circumvent the most difficult task, i.e.,
expanding the three-body phase space integral.
This procedure turns out to be the simplest in practice,
especially when contrasted with the orthodox matching method
outlined in section~\ref{matching:stragey:shape:function}.
This will be exemplified by more concrete
examples in section~\ref{Description:matching:calculation}.

In (\ref{F2:definition}), we have refrained from expressing $F_2$ by
taking the second-order derivatives of the quark amplitude over $q$,
as frequently adopted in many works~\cite{Bodwin:2003wh,He:2007te,Sang:2009,Fan:2009zq}.
The reason is that we try to avoid potential ambiguity associated with this operation,
which usually happens when one performs the standard expansion around $m_c$.
The recipe given in (\ref{F2:definition}) is unambiguous and simple provided
that $E_q$ is kept fixed.
The expression obtained this way are connected to the standard one through
reshuffling some terms between $F_0$ and $F_2$.

Squaring the matrix element (\ref{ampl:ccbar}), summing over final-state polarizations
and averaging upon the initial-state spins, we obtain
\bqa
\overline{\sum} \, \left| {\mathcal M}[c\bar{c}(^3S_1,\lambda)+gg]
\right|^2 &=&
 4 E_q^2 \overline{\sum} \, \left\{|F_0(\lambda)|^2 \langle
{\mathcal O}^{c\bar{c}}_1 \rangle + 2\,{\rm Re}[F_0 F_2^*] \langle
{\mathcal P}^{c\bar{c}}_1 \rangle +\cdots \right\}
\nn \\
&=&  4 E_q^2 (2N_c) \overline{\sum} \, \left\{|F_0(\lambda)|^2
 + 2\,{\rm Re}[F_0 F_2^*]  {{\mathbf q}^2\over m_c^2} + \cdots
\right\}.
\label{ccbar:ampl:squared}
\eqa
The matrix elements $\langle{\mathcal O}_1^{\,c\bar{c}}\rangle$
and $\langle {\mathcal P}_1^{\,c\bar{c}}\rangle$ denote the vacuum
expectation values of the production operators for producing
the free $c\bar{c}({}^3S_1^{(1)})$ state, which are given by
\begin{subequations}
\bqa
\langle {\mathcal O}_1^{\,c\bar{c}}\rangle  &=& 2 N_c,
\label{O1ccbar:vev}%
\\
\langle {\mathcal P}_1^{c\bar{c}} \rangle  &=&  {{\mathbf q}^2\over
m_c^2}\langle {\mathcal O}_1^{\,c\bar{c}}\rangle,
\label{P1ccbar:vev}%
\eqa
\end{subequations}
where the factor of $2N_c$ in the right side of
Eq.~(\ref{O1ccbar:vev}) arises from the spin and color normalization
factors for free $c\bar{c}$ states.
Comparing both sides of (\ref{ccbar:ampl:squared}), one may directly
deduce the short-distance coefficients $|F_0|^2$ and ${\rm Re}[F_0 F_2^*]$.

In passing it may be worth reminding that,
during the polarization sum/average procedure, new factors of $E_q$ will be
unavoidably regenerated in the squared amplitude.
Evidently, such factors can arise from summing the
polarization states of $c\bar{c}({}^3S_1^{(1)})$ state. In the
standard way of expansion, one needs re-expand these occurring
$E_q$ factors once and more, and keeping reshuffling the corresponding terms from the
LO matrix element squared to the relativistic
correction piece. Fortunately, since we keep $E_q$ fixed in our approach,
no any extra labor needs to be invested for such complication.
This comprises another attractive trait of our expansion strategy.

Substituting the short-distance coefficients
(\ref{F0:and:F2:definition}) to (\ref{ampl:squared:jpsi:gg}),
or directly converting
the quark amplitude squared (\ref{ccbar:ampl:squared}) to (\ref{ampl:squared:jpsi:gg}),
after some straightforward algebra,
one then obtains the desired squared matrix element for producing $J/\psi$
plus light hadrons.

There arises one immediate question. Since $m_c$ has been eliminated
in favor of $E_q$
in the physical matrix element squared, it is necessary to specify
which value of $E_q$ should be taken, in order to make concrete predictions.
If there were no rationale to restrict its value, our approach would just
yield {\it ad hoc} predictions and lack attractiveness.

Fortunately, the answer to this question is definite, i.e.,
theoretical consistency requires that $E_q$ can be fixed in an
unambiguous manner. To see this, let us first make the
following observation:
\begin{subequations}
\bqa
2 E_q &=&  2m_c \left( 1 + {1\over 2} {{\mathbf q}^2\over m_c^2}+
O({\mathbf q}^4)\right),
\label{einstein:eq}%
\\
M_{J/\psi} &=& 2 m_c \left(  1+ {1\over 2} \langle v^2
\rangle_{J/\psi}  + O(v^4) \right),
\label{G-K:rel}%
\eqa
\end{subequations}
where (\ref{einstein:eq}) comes from simple nonrelativistic kinematics,
the inverse relation  of (\ref{expand:m:in:Eq}).
Eq.~(\ref{G-K:rel}) is nothing but the G-K relation
(\ref{G-K:relation:first}).

The similarity between (\ref{einstein:eq}) and (\ref{G-K:rel})
strongly suggests that, $E_q$ appearing
everywhere in the short-distance coefficients
in (\ref{ampl:squared:jpsi:gg})
can be replaced by $M_{J/\psi}/2$. Naively, the entering of $J/\psi$ mass
into short-distance coefficients seems to be a nuisance, which is against
the doctrine of the EFT. Nevertheless, this procedure is valid,
at least to the present accuracy of order $v^2$,
thanks to the G-K relation~\footnote{
It is worth noting that this substitution
has also been adopted in a recent investigation of the
relativistic corrections to the exclusive charmonium production process
$e^+e^-\to J/\psi+\eta_c$~\cite{Bodwin:2007ga}.}.

Identification of $2E_q$ with $M_{J/\psi}$, in conjunction
with our nonstandard expansion (\ref{expand:m:in:Eq}),
turn out to have great advantages.
By this way, the relativistic effects in phase space integrals
are automatically incorporated. In some sense,
our approach fulfills the role of the color-singlet shape-function
by promoting the partonic kinematics to hadronic kinematics,
but not necessarily restricted to the region of the maximum $J/\psi$ energy.
In addition, since the mass of $J/\psi$ is known rather precisely,
it is better to choose it as the input parameter than the
ambiguously defined charm quark mass.
To summarize, our method greatly simplifies the efforts required for the
matching calculation, by reducing the task of matching the cross section to
matching the amplitude squared.

\section{Three-body phase space for
$\bm{e^+}\bm{e^-}  \bm{\to} \bm{c}\bar{\bm{c}}\bm{({}^3S_1)}\bm{+}\bm{gg}$
\label{3:body:phase:space}}

One integral part of the matching procedure is to consistently
incorporate the relativistic effects inherent in the phase space
integration. As was explained in section~\ref{matching:stragey:mine}, owing to
the virtue of our matching approach,
no special care needs to be paid to the phase space integral,
provided that we identify the invariant mass of $c\bar{c}$ pair, $2E_q$,
to be $M_{J/\psi}$.
For the process $e^-(l_1)+e^+(l_2)\to \gamma^\ast(K)\to
c\bar{c}({}^3S_1^{(1)},P)+g(k_1)+g(k_2)$ considered in this work,
the energy-momentum conservation requires
$K=l_1+l_2=P+k_1+k_2$. The electron and gluon are treated
to be massless, and $P^2= M_{J/\psi}^2$.
We will evaluate the three-body phase space $d\Pi_3$ in the $e^+ e^-$
center-of-momentum (laboratory) frame.

The center-of-mass energy squared is defined by $s \equiv K^2$. It is also
convenient to define a dimensionless ratio $r \equiv M^2_{J/\psi}/
s$. The differential three-body phase space can be expressed as follows:
\bqa
\int \!\! d \Pi_3 &=&  \int \!\! {d^3 P \over (2\pi)^{3}2P^0}
{d^{3}k_1 \over (2\pi)^{3}2k_1^0} {d^{3}k_2 \over (2\pi)^{3}2k_2^0}
(2\pi)^{4} \delta^{(4)}(K-P-k_1-k_2)
\nn\\
&=& {s\over 2(4\pi)^4}  \int_{2\sqrt{r}}^{1+r} \!\! dz \int^1_{-1}
\!\! d \cos \theta \int^{x_1^+}_{x_1^-} \!\! dx_1 \int^{2\pi}_0 \!\!
d \Phi^*_1.
\label{PS3-def-4-variables}
\eqa
It is worth noting that, in contrast to (\ref{phase:space:orthdox:matching}),
the advantage of our matching method is there is no need to expand $P$
around the momentum of a fictitious particle with mass of $2m_c$.
For notational simplicity, we have introduced three dimensionless
variables, $z \equiv {2 P\cdot K\over K^2}= {2 P^0\over\sqrt{s}}$,
$x_1 \equiv {2 k_1 \cdot K\over K^2}= {2 k_1^0\over\sqrt{s}}$,
$x_2 \equiv {2 k_2 \cdot K\over K^2}= {2 k_2^0\over\sqrt{s}}$, to
characterize the fractional energies
carried by the $c\bar{c}({}^3S_1^{(1)})$ pair, the gluon 1, and
gluon 2 in the laboratory frame.
Only two of these three variables are independent, since they are subject to
the constraint from energy conservation: $x_1+x_2+z=2$.
In the following, we will eliminate $x_2$ everywhere
in favor of $z$ and $x_1$ as the integration variables.

We use ($\theta$, $\phi$) to denote the polar and azimuthal angles of
the outgoing $J/\psi$ momentum with respect to the moving direction
of $e^-$ in the laboratory frame. We have suppressed $d \phi$ in the
integration measure since it has been trivially integrated over
due to the axial symmetry of the reaction under consideration.
It is convenient to introduce a set of auxiliary solid-angle variables
$(\Theta_1^*,\Phi^*_1)$ as the polar and azimuthal angles of the
moving direction of the gluon 1 in a rotated coordinate system
relative to the laboratory frame, where the $J/\psi$ moves along
the new $+\hat{z}$ axis. Given the energy fractions $z$ and $x_1$,
four-momentum conservation uniquely constrains the polar angle
$\Theta_1^*$:
\beq
\cos\Theta_1^* = {2(1+r-z)-x_1(2-z) \over x_1 \sqrt{z^2-4r}}.
\label{Big:Theta_1*}%
\eeq

The main advantage of choosing these integration variables as given in
(\ref{PS3-def-4-variables}), is that each of them has an
intuitive interpretation and the respective integration boundaries are
rather simple. This is in contrast with the set of variables employed
in Ref.~\cite{Cho:1996cg,He:2007te,He:2009uf}.

The integration boundaries for $z$ have been explicitly labeled in
(\ref{PS3-def-4-variables}), and those for $x_1$ can be easily
inferred:
\bqa
x_1^\pm &=& {2-z \pm \sqrt{z^2-4r} \over 2}.
\label{x1:int:boundary}
\eqa
For the suppressed variable $x_2$, the boundaries would be
the exactly same as $x_1$.

If one concentrates only on the energy spectrum of (un)polarized $J/\psi$,
disregarding its angular distribution, one may take a shortcut-- by
starting with the simpler process $\gamma^\ast\to J/\psi+gg$, then
converting the differential decay rate
to the $J/\psi$ differential production cross section~\cite{Keung:1980ev}.
In such case, since there is no preferred orientation in space,
two angular variables, $\cos\theta$ and $\Phi^*_1$,
can be trivially integrated over in (\ref{PS3-def-4-variables}),
consequently one is left with only two dimensionless energy variables in the
three-body-phase-space measure:
\bqa
\int \!\! d \Pi_3 &=&  {s\over 2(4\pi)^3} \int_{2\sqrt{r}}^{1+r} dz
\int^{x_1^+}_{x_1^-} dx_1.
\label{PS3-def-2-variables}%
\eqa

In some situation, it is desirable to know the analytic expression for
the integrated $J/\psi$ production rate.
To this purpose, it seems more advantageous to choose a different order to
perform the phase-space integration:
\bqa
\int \!\! d \Pi_3 &=&  {s\over 2(4\pi)^3} \int_{0}^{1-r} d x_1
\int^{1+r}_{1-x_1+{r\over 1-x_1}} d z.
\label{PS3-def-2-variables:altnative}%
\eqa
As an intermediate byproduct, one can deduce the gluon energy spectrum once performing
the integration over $z$. Phenomenologically, knowing this is not so meaningful,
since it cannot be directly linked with a physical observable.
However, as a calculational device, choosing this particular order for phase-space integration
leads to considerable technical simplicity, because the integration boundaries in
(\ref{PS3-def-2-variables:altnative})
are far simpler than those in (\ref{PS3-def-2-variables}).
As a consequence, by this way one can readily
deduce the $J/\psi$ total production rate in a closed form, which is otherwise rather difficult
to achieve if one starts from (\ref{PS3-def-2-variables}).

\section{Outline of matching calculations for
$\bm{e^+ e^- \to} \bm{c}\bar{\bm{c}} \bm{({}^3S_1)} \bm{+} \bm{gg}$
\label{Description:matching:calculation}}

In this section, we present a detailed description on how the
short-distance coefficients through relative order-$v^2$
can be determined via our matching
procedure, concretizing the method put forward in
section~\ref{matching:stragey:mine}. We also illustrate how the
various types of predictions for the inclusive $J/\psi$ production rate in
$e^+e^-$ annihilation emerge.

\begin{figure}[!htb]
\centerline{
\includegraphics[width=8.0cm]{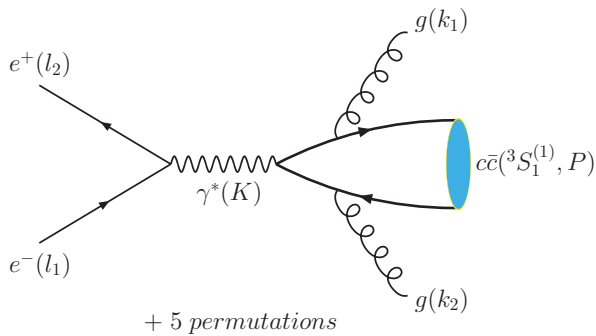}
} \caption{Lowest-order Feynman diagrams for $e^+ e^-\to
J/\psi+gg$.} \label{Feynman:diag}
\end{figure}

In order to deduce the intended short-distance coefficients, one
needs to consider the parton process $e^+e^-\to
c\bar{c}({}^3S_1^{(1)},P,\lambda)+gg$, with one typical lowest-order
diagram shown in Fig.~\ref{Feynman:diag}. The calculation is
expedited by the covariant projection technique developed
by Bodwin and Petrelli~\cite{Bodwin:2002hg}, which helps to
readily project out the amplitude for the $c\bar{c}$
pair being in the color-singlet spin-triplet state~\footnote{An
alternative approach, the {\it threshold expansion
method}~\cite{Braaten:1996jt}, is also valid to deduce the NRQCD
matching coefficients. This method has been utilized to investigate
the order-$v^2$ relativistic correction to $J/\psi$ photoproduction at
HERA~\cite{Paranavitane:2000if}. However, for the process
with involved kinematics like ours, this method, which requires to
extensively deal with the algebra of two-component spinors,
seems not as convenient as the covariant projection technique outlined
in \cite{Bodwin:2002hg}.}.

Our matching method will be exemplified by the following three
subsections. In section~\ref{matching:unpol:Jpsi:distr} and
section~\ref{matching:long-pol:Jpsi:distr}, where we are only
interested in the energy spectra of unpolarized and
longitudinally-polarized $J/\psi$, we take the shortcut by
considering the simpler process $\gamma^\ast \to
c\bar{c}({}^3S_1^{(1)},P,\lambda)+gg$; in
section~\ref{matching:unpol:Jpsi:ang:distr:e+e-}, where we are also
interested in the angular as well as the energy distributions of
$J/\psi$, we work with the full process $e^+e^-\to
c\bar{c}({}^3S_1^{(1)},P,\lambda)+gg$.

\subsection{Matching calculation for $\bm{\gamma^\ast \to}
\bm{c}\bar{\bm{c}} \bm{({}^3S_1)} \bm{+} \bm{gg}$ %
\label{matching:unpol:Jpsi:distr}}

We start with the tree-level quark amplitude $\gamma^\ast(K)\to c(p)
\bar{c}(\bar{p})+ g(k_1) + g(k_2)$, with the momenta of $c$ and
$\bar{c}$ defined in (\ref{p-pbar-mom})~\footnote{The calculation presented
in this subsection is somewhat similar to that for the process $g^*\to J/\psi+gg$,
from which
the relativistic correction to the gluon-to-$J/\psi$ color-singlet fragmentation function
can be extracted~\cite{Bodwin:2003wh}.}.
The amplitude can be written as
\begin{equation}
\bar{u}(p) \mathscr{A}  v(\bar{p}) = \textrm{Tr} \big[ v(\bar{p})
\bar{u}(p) \mathscr{A} \big].
\label{amplitud:ccbar}%
\end{equation}
Here $\mathscr{A}$ is a Dirac-color-space matrix, which reads
\beq
\mathscr{A} = (e_c e g_s^2) \, T^a T^b \otimes \, /\!\!\!
\epsilon^\ast(g;k_1)  {1\over /\!\!\!p+/\!\!\!k_1-m_c} /\!\!\!
\epsilon(\gamma^*;K)  {1\over -/\!\!\!\bar{p}-/\!\!\!k_2-m_c}
 /\!\!\!
\epsilon^\ast(g;k_2) + 5\;\;\textrm{perms},
\label{A:concrete:form}%
\eeq
where $g_s$ signifies the QCD coupling strength, and $e e_c$
denotes the electric charge of the charm quark ($e_c={2 \over 3}$).
$a$, $b$ denote the color indices of the two gluons,
$\epsilon(\gamma^*;K)$, $\epsilon^\ast(g;k_1)$ and
$\epsilon^\ast(g;k_2)$ represent the polarization vectors of the
decaying virtual photon, gluon 1 and 2, respectively.

One of the important sources of relativistic corrections stem from
expanding the quark propagators. Apart from retaining the factor $q$
in the numerator of the propagator, one needs also expand its
denominator to the quadratic order in $q$. Taking the diagram shown
in Fig.~\ref{Feynman:diag} as an example, two propagators there are
expanded to be
\begin{subequations}
\bqa
{1\over (p+k_1)^2-m^2_c}&=& {1\over P\cdot k_1}-{2 q\cdot k_1\over
(P\cdot k_1)^2} + {4 (q\cdot k_1)^2 \over (P\cdot k_1)^3}+O(q^3),
\label{fig1:propag:1:expand}
\\%
{1\over (\bar{p}+k_2)^2-m^2_c}&=& {1\over P\cdot k_2} + {2 q\cdot
k_2\over (P\cdot k_2)^2} + {4 (q\cdot k_2)^2 \over (P\cdot
k_2)^3}+O(q^3).
\label{fig1:propag:2:expand}
\eqa
\end{subequations}

To proceed, we need project the amplitude (\ref{amplitud:ccbar})
onto the spin-triplet color-singlet $c(p)\bar{c}(\bar{p})$ state, by
replacing the $v(\bar{p})\bar{u}(p)$ with a suitable projection
matrix. The projector that is valid to all orders in $\mathbf{q}$
for the spin-triplet color-singlet channel, denoted by
$\Lambda_3^{(1)}(p,\bar{p},\lambda)$ ($\lambda$ characterizes the
polarization of this spin-triplet $c\bar{c}$ pair), assumes the
particular form~\cite{Bodwin:2002hg}:
\bqa
\Lambda_3^{(1)}(p,\bar{p},\lambda)&=&
 -\frac{1}{4\sqrt{2}E_q (E_q + m_c\,)} (/\!\!\!\bar{p}-m_c)
\,/\!\!\!\epsilon^{\,\ast}(\lambda) (\,/\!\!\!\!P\!+\!2E_q )
(/\!\!\!{p}+m_c)\otimes  {\mathbf{1}_c\over \sqrt{N_c}},
\label{spin-projector}%
\eqa
where $\mathbf{1}_c$ is the unit matrix in the fundamental
representation of the color $SU(3)$ group, and the spin-polarization
vector $\epsilon^\ast(\lambda)$ satisfies $P \cdot
\epsilon^\ast(\lambda)=0$.  The above spin projector is derived by
assuming the relativistic normalization convention for Dirac spinor:
$\bar u^{(r)} u^{(s)}=2 m \delta_{rs}$ and $ u^{(r)\dagger}
u^{(s)}=2 E_q \delta_{rs}$. Applying this spin-color projector to
(\ref{amplitud:ccbar}), we obtain
\beq
\mathcal{M}_{c\bar{c}}(P,q,\lambda;k_1,k_2) =
\textrm{Tr}\big\{\mathscr{A} \, \Lambda_3^{(1)}(p,\bar{p},\lambda)
\big\},
\label{amp-spin-double}%
\eeq
where the trace acts on both Dirac and color spaces.
$\mathcal{M}_{c\bar{c}}(P,q,\lambda;k_1,k_2)$ can be interpreted as
the amplitude for producing a color-singlet, spin-triplet $ c\bar{c}$
pair in association with two gluons.

We have emphasized that our method differs from the conventional tenet of
matching. Rather than Taylor-expanding $E_q$ around $m_c$
in ${\mathbf q^2/m_c^2}$ everywhere in $\mathcal{M}_{c\bar{c}}(P,q,\lambda;k_1,k_2)$,
we choose to expand $m_c$ around $E_q$ in powers of
${\mathbf q^2/E_q^2}$ using (\ref{expand:m:in:Eq}). It is worth reminding that
we should not forget to trade $m_c$ for $E_q$ that appears in the
denominator of the projector (\ref{spin-projector}):
\beq
{1\over E_q+m_c} = {1\over 2 E_q} \left( 1 + {1\over 4} {{\mathbf
q}^2\over E_q^2}+O({\mathbf q}^4) \right).
\label{exp:projector:denom}%
\eeq
It is convenient, at this stage, to truncate the amplitude
$\mathcal{M}_{c\bar{c}}$ such that all the terms in it are at most
quadratic in $q$.

In the amplitude (\ref{amp-spin-double}), the
$c\bar c$ pair is warranted to be in the spin-triplet, but not
necessarily in the $S$-wave orbital-angular-momentum state. To
project out the $S$-wave amplitude, one needs average the amplitude
$\mathcal{M}_{c\bar{c}}$ over all the direction of the relative
momentum $\mathbf{q}$ in the rest frame of $c\bar{c}(P)$ pair.
This literal angular averaging procedure can help to
acquire a specific class of relativistic corrections to all orders in $v$.
However it is feasible only for few processes with very simple
kinematics~\cite{Bodwin:2002hg,Bodwin:2007ga}.
For the process at hand, this procedure would become extremely
cumbersome, if not impossible.
Fortunately, to the intended ${\mathcal O}(v^2)$
accuracy, one can utilize a standard trick to project out the $S$-wave
part. Now we already have the amplitude truncated up to two powers of $q$.
Terms that contain no powers of $q$,
or contain explicitly the Lorentz scalar $q^2$
(which can be translated into $-{\mathbf q}^2$),
already yield a pure $S$-wave contribution;
for those terms containing the tensor $q^\mu
q^\nu$ contracted with other 4-vectors, we can make the following
substitution to extract the $S$-wave piece~\footnote{Note
this $S$-wave projection operation will regenerate
factors of $E_q$ in the denominator of the amplitude. Obviously it
will not cause any trouble for our strategy of matching. In
contrast, in the orthodox matching recipe, one is forced to reexpand
those terms containing these newly-generated $E_q$ factors, and
reshuffling the corresponding terms from the ``leading-order" piece
to the relativistic correction piece. This further exhibits the
merit of our method.}:
\bqa
q^\mu q^\nu  &\to&  {\mathbf{q}^2 \over 3}\: \Pi^{\mu\nu}(P),
\label{amp-QQ-angle}%
\eqa
where
\bqa
\Pi^{\mu\nu}(P) &\equiv& -g^{\mu\nu}+{P^\mu P^\nu\over P^2}.
\label{Pi:spin:1} %
\eqa

Following all these steps, we then obtain the desired ${}^3S_1$ piece of the
amplitude, ${\mathcal M}({}^3S_1,P,\lambda;k_1,k_2)$,
accurate through the order ${\mathbf q}^2$.
Following what has been elaborated in section~\ref{matching:stragey:mine},
at this stage it is legitimate to replace $E_q$ everywhere by $M_{J/\psi}/2$
in ${\mathcal M}({}^3S_1,P,\lambda;k_1,k_2)$.

It is now the time to deduce the desired ``short-distance"
coefficients $F_0$ and $F_2$, following the recipes given in
(\ref{F0:definition}) and (\ref{F2:definition}). After this is done,
we need square these coefficients and perform the corresponding
spin-color sum/average:
\begin{subequations}
\bqa
\overline{\sum} \, \left| F_0 \right|^2 & = & {1\over 3}\,
\Pi_{\mu\mu^\prime}(K)\,
\Pi_{\rho\rho^\prime}(P)\,(-g_{\alpha\alpha^\prime})(-g_{\beta\beta^\prime})\,
{\mathcal F}_0^{\mu;\; \rho \alpha\beta} {\mathcal
F}_0^{\,\ast\,\mu^\prime;\; \rho^\prime \alpha^\prime \beta^\prime},
\label{unpol:Jpsi:square:F0}
\\%
2\,\overline{\sum}  \, {\rm Re} \left[ F_0 F_2^\ast \right]
& = &  {1\over 3}\,\Pi_{\mu\mu^\prime}(K)\,
\Pi_{\rho\rho^\prime}(P)\,(-g_{\alpha\alpha^\prime})
(-g_{\beta\beta^\prime})\,2\,{\rm Re}\big[ {\mathcal
F}_0^{\mu;\;\rho \alpha\beta}  {\mathcal
F}_2^{\,\ast\,\mu^\prime;\,\rho^\prime \alpha^\prime
\beta^\prime}\big],
\label{unpol:Jpsi:F0:F2}
\eqa
\label{unpol:Jpsi:F-i}
\end{subequations}
where the amputated ``short-distance" coefficients ${\mathcal F}_i$
($i=0,2$) are defined through
\bqa
F_i &=& {\mathcal F}_i^{\mu;\;\rho \alpha\beta}
\epsilon_\mu(\gamma^\ast;K)
\,\epsilon^\ast_\rho({}^3S_1;P,\lambda)\,
\epsilon^\ast_\alpha(g;k_1)\,\epsilon^\ast_\beta(g;k_2).
\eqa
For simplicity, we have suppressed the color indices,
implicitly contained in ${\mathcal F}_i$, which reads ${\rm Tr}(T^a
T^b)/\sqrt{N_c}$. In Eqs.~(\ref{unpol:Jpsi:F-i}), we have summed over polarization and color
of the $c\bar{c}(^3S_1^{(1)})$ pair and two gluons, and averaged
upon three spin states of the virtual photon (note the
prefactor ${1\over 3}$). Polarization sum for
two gluon states has been taken into account by the metric tensors
$-g_{\alpha\alpha^\prime}$ and $-g_{\beta\beta^\prime}$, and that
for the virtual photon and $J/\psi$ by the polarization tensor
$\Pi^{\mu\mu^\prime}(K)$ and $\Pi^{\rho\rho^\prime}(P)$.

According to equation (\ref{ampl:squared:jpsi:gg}), we then obtain
the squared matrix element for $\gamma^\ast\to J/\psi+gg$. Including
the 3-body phase space measure (\ref{PS3-def-2-variables}), we can
obtain the differential decay rate for unpolarized $J/\psi$:
\bqa
{d \Gamma[\gamma^\ast\to J/\psi+gg]\over dz dx_1 }&=& {1\over 2!}
{\sqrt{s} \over 4(4\pi)^3}  \overline{\sum} \, \left|{\mathcal
M}\big[\gamma^\ast\to J/\psi+gg\big] \right|^2,
\label{diff:decay:rate:unpol:Jpsi}%
\eqa
where we have included a statistical factor of ${1\over 2!}$ to
account for the indistinguishability of two gluons.

To convert the differential decay rate in
Eq.~(\ref{diff:decay:rate:unpol:Jpsi}) into $J/\psi$ production
cross section, one can use the formula~\cite{Keung:1980ev}
\bqa
{d \sigma[e^+ e^-\to J/\psi+gg]\over dz d x_1} &=& {4\pi\alpha\over
s^{3/2}}\,{d \Gamma[\gamma^\ast\to J/\psi+gg]\over dz d x_1}.
\label{diff:cross:section:unpol:Jpsi}%
\eqa
It is not difficult to integrate over the fractional energy of
gluon 1, $x_1$, with the integration boundaries specified in
(\ref{x1:int:boundary}) to acquire the energy spectrum for unpolarized
$J/\psi$.

\subsection{Matching calculation for $\bm{\gamma^\ast \to}
\bm{c}\bar{\bm{c}} \bm{({}^3S_1, \lambda=0)} \bm{+} \bm{gg}$%
\label{matching:long-pol:Jpsi:distr}}

It is also interesting to know how the polarization
information of $J/\psi$ varies with its energy.
To accomplish this, in addition to the
differential energy spectrum for unpolarized $J/\psi$ as given in
section~\ref{matching:unpol:Jpsi:distr}, we also need know that
for the longitudinally-polarized $J/\psi$. In this section
we outline how the corresponding matching calculation is
carried out.

The ``short-distance" coefficients $F_0$ and $F_2$ can be obtained
following the the same procedure as outlined in
section~\ref{matching:unpol:Jpsi:distr}. Nevertheless, the
longitudinal polarization vector of $J/\psi$ can be explicitly
substituted by
\beq
\epsilon_L^{\ast\,\mu}({}^3S_1) \equiv
\epsilon_L^{\ast\,\mu}({}^3S_1;P,\lambda=0) = {P^0\over 2 E_q
|{\mathbf P}|} P^\mu - {2 E_q \over \sqrt{s} \, |{\mathbf P}|}
K^\mu,
\label{pol:vec:long}%
\eeq
which satisfies $\epsilon_L \cdot P=0$ and
$\epsilon_L \cdot \epsilon_L^\ast =-1$.

We then square these coefficients and perform the corresponding
spin-color sum/average:
\begin{subequations}
\bqa
& &\overline{\sum} \, \left| F_0 \right|^2 =  {1\over 3} \,
\Pi_{\mu\mu^\prime}(K)\,\epsilon_{L\,\rho}^{\ast}({}^3S_1)\,
\epsilon_{L\,\rho^\prime}({}^3S_1)\,
(-g_{\alpha\alpha^\prime})(-g_{\beta\beta^\prime})\, {\mathcal
F}_0^{\mu;\; \rho \alpha\beta} {\mathcal F}_0^{\,\ast\,\mu^\prime;\;
\rho^\prime \alpha^\prime \beta^\prime},
\\%
&& 2\,\overline{\sum}  \, {\rm Re} \left[ F_0 F_2^\ast \right]
=  {1\over 3}
\,\Pi_{\mu\mu^\prime}(K)\,\epsilon_{L\,\rho}^{\ast}({}^3S_1)\,
\epsilon_{L\,\rho^\prime}({}^3S_1)\, (-g_{\alpha\alpha^\prime})
(-g_{\beta\beta^\prime})\,2\,{\rm Re}\big[ {\mathcal
F}_0^{\mu;\;\rho \alpha\beta}  {\mathcal
F}_2^{\,\ast\,\mu^\prime;\,\rho^\prime \alpha^\prime
\beta^\prime}\big].
\nn\\
\label{}
\eqa
\end{subequations}
Here the amputated ``short-distance" coefficients $\mathcal{F}_i$
are the same as what appear in Eqs.~(\ref{unpol:Jpsi:F-i}).
Obviously the sum over polarizations needs
not act on the $c\bar{c}({}^3S_1,\lambda=0)$ state.

Apart from replacing $E_q$ everywhere with $M_{J/\psi}/2$,
we also need substitute $P^0={\sqrt{s}\over 2}z$ and $|{\mathbf P}|= {\sqrt{s}\over
2}\sqrt{z^2-4 r}$ for $\epsilon_L^\ast$ in above expressions. We
then follow (\ref{ampl:squared:jpsi:gg}) to obtain the squared
matrix element for $\gamma^\ast\to J/\psi_L+gg$. With this
expression at hand, one can use
(\ref{diff:decay:rate:unpol:Jpsi}) to infer the differential decay rate
from a virtual photon, subsequently use
(\ref{diff:cross:section:unpol:Jpsi}) to deduce the corresponding
differential cross section for producing the longitudinal-polarized
$J/\psi$ in $e^+e^-$ annihilation.

\subsection{Matching calculation for $\bm{e^+ e^- \to}
\bm{c}\bar{\bm{c}} \bm{({}^3S_1)} \bm{+} \bm{gg}$
\label{matching:unpol:Jpsi:ang:distr:e+e-}}

In section~\ref{matching:unpol:Jpsi:distr} and \ref{matching:long-pol:Jpsi:distr},
we have resorted to a shortcut by considering the production rate of a $J/\psi+gg$
from a virtual photon decay, since the sole purpose is to deduce the
energy spectrum of unpolarized or longitudinally-polarized $J/\psi$ in $e^+e^-$ annihilation.
In this subsection, we are interested in knowing the angular-energy
double differential distribution of unpolarized $J/\psi$.
To this end, it is compulsory to begin with the full
process $e^-(l_1)e^+(l_2)\to J/\psi(P)+g(k_1)g(k_2)$. We use $l_1$
and $l_2$ to signify the momenta of $e^-$ and $e^+$, respectively,
and $l_1+l_2=K$.

The main result derived in section~\ref{matching:unpol:Jpsi:distr}
can be directly transplanted here. In particular, the
``short-distance" coefficients $F_0$ and $F_2$ for $\gamma^\ast\to
J/\psi+gg$, determined there by employing (\ref{F0:definition}) and
(\ref{F2:definition}), only need undergo some slight modifications
to meet our purpose. That is, one needs replace the polarization
vector of the virtual photon by a $e^+ e^-$ bispinor and insert a
photon propagator and a QED coupling.
These slight changes are embodied in squaring
these coefficients and performing the corresponding spin-color
sum/average:
\begin{subequations}
\bqa
\overline{\sum} \, \left| F_0 \right|^2 & = & {1\over 4} \,
L_{\mu\mu^\prime}{e^2\over s^2}\, \Pi_{\rho\rho^\prime}(P)
\,(-g_{\alpha\alpha^\prime}) (-g_{\beta\beta^\prime})\, {\mathcal
F}_0^{\mu;\; \rho \alpha\beta} {\mathcal F}_0^{\,\ast\,\mu^\prime;\;
\rho^\prime \alpha^\prime \beta^\prime},
\label{e+e-:unpol:Jpsi:square:F0}
\\%
2\,\overline{\sum}  \, {\rm Re} \left[ F_0 F_2^\ast \right]
& = &  {1\over 4}\, L_{\mu\mu^\prime}{e^2\over
s^2}\,\Pi_{\rho\rho^\prime}(P)\,(-g_{\alpha\alpha^\prime})
(-g_{\beta\beta^\prime})\,2\,{\rm Re}\big[ {\mathcal
F}_0^{\mu;\;\rho \alpha\beta}  {\mathcal
F}_2^{\,\ast\,\mu^\prime;\,\rho^\prime \alpha^\prime
\beta^\prime}\big].
\label{e+e-:unpol:Jpsi:F0:F2}
\eqa
\label{e+e-:unpol:Jpsi:F-i}
\end{subequations}
Here the amputated ``short-distance" coefficients $\mathcal{F}_i$
are the same as what appear in Eqs.~(\ref{unpol:Jpsi:F-i}).
The factor $1/4$ represents the average
over polarizations of the initial $e^-e^+$ state,
and $L^{\mu\mu^\prime}$ denotes the leptonic tensor:
\bqa
L^{\mu\,\mu^\prime} &=& \sum_{s,r} \big[\bar{v}(l_2;r)\gamma^\mu
u(l_1;s)\big] \big[\bar{u}(l_1;s)\gamma^{\mu^\prime} v(l_2;r)\big]
\nn\\
 & = & 4 \,\big( l_1^\mu \,l_2^{\mu^\prime} + l_2^\mu\,l_1^{\mu^\prime}-
l_1\cdot l_2 \,g^{\mu\mu^\prime} \big),
\label{leptonic:tensor}%
\eqa
where the sum is extended over all possible polarization states of
the electron and positron.

Substituting Eqs.~(\ref{e+e-:unpol:Jpsi:F-i}) into (\ref{ampl:squared:jpsi:gg}), we
then obtain the color-spin averaged/summed matrix element squared
for the process $e^+ e^-\to J/\psi+gg$. Including the 3-body phase
space measure (\ref{PS3-def-4-variables}) and the flux factor, we
can obtain the differential production rate for unpolarized $J/\psi$:
\bqa
{d \sigma \big[e^+ e^-\to J/\psi+gg \big]\over dz\, d \cos \theta\,
dx_1\,d\Phi^*_1} &=&  {1\over 2!} {1\over 4(4\pi)^4}  \,
\overline{\sum} \, \left|{\mathcal M} \big[e^+ e^-\to J/\psi+gg\big]
\right|^2.
\label{diff:Xsection:jpsi:e+e-}
\eqa

In the squared matrix elements, all the scalar products can be
expressed in terms of $z$, $\theta$, $x_1$, and one additional angular
variable, $\theta_1$, which represents the angle between the
3-momentum of gluon 1 and the beam direction in the laboratory
frame. This polar angle is connected to $\theta$, $\Theta_1^*$ and $\Phi_1^*$
through
\bqa
\cos \theta_1 &=& \cos\theta \cos\Theta_1^* -\sin\theta \sin
\Theta_1^* \cos\Phi_1^*,
\eqa
where $\cos\Theta_1^*$ is uniquely determined when $z$ and $x_1$ are
given, as indicated in (\ref{Big:Theta_1*}).

As first elaborated in \cite{Cho:1996cg}, general consideration
based on Lorentz invariance, parity and gauge invariance demands that
for inclusive $J/\psi$ production in $e^+e^-$ annihilation,
the double differential distribution must bear the following
form:
\bqa
{d \sigma \big[e^+ e^-\to J/\psi + X \big] \over dz d \cos\theta}
&=& S(z)[1+A(z) \cos^2\theta].
\label{double:diff:distr:form}
\eqa

It is interesting to note that,
after the suitable reduction, the squared matrix element for this reaction
can depend upon the polar angles only through rather limited
combinations-- 1, $\cos^2\theta$, $\cos\theta_1\cos\theta$,
and $\cos^2\theta_1$, respectively.
Therefore, to arrive at the expression indicated
in (\ref{double:diff:distr:form}),
suffices it to know the following integrations over $d\Phi_1^*$:
\begin{subequations}
\bqa
\int^{2\pi}_0 d\Phi_1^*  &=& 2\pi,
\label{int:costheta1:0}
\\
\int^{2\pi}_0 d\Phi_1^* \cos \theta_1 &=& 2\pi \cos\theta
\cos\Theta_1^*,
\label{int:costheta1:1}
\\
\int^{2\pi}_0 d\Phi_1^* \cos \theta^2_1 &=& 2\pi \left[{1\over 4}
\left(1+\cos^2 \theta \right) + \left({1\over 4}-{1\over 2}
\cos^2\Theta_1^*\right) (1-3\cos^2\theta)\right].
\label{int:costheta1:2}
\eqa
\end{subequations}
Thus we are reassured that only the zeroth and second powers of
$\cos\theta$ are allowed to appear in the
double differential distribution for $J/\psi$ production,
in conformity to (\ref{double:diff:distr:form}).
It may be also worth pointing out that, one great simplification can be
made insofar as only the differential energy spectrum of unpolarized
$J/\psi$ is concerned. In this case, the second term inside the square
bracket in (\ref{int:costheta1:2}) can be discarded, since its contribution
vanishes upon integration over $\theta$.

\section{Inclusive $\bm{J/\psi}$ distributions at $\bm{B}$ factory}
\label{phenomena:Jpsi:distr:B:factory}

In this section, we report our results of order-$v^2$ relativistic correction
to inclusive $J/\psi$ production associated with non-$c\bar{c}$ states at $B$ factory
at $\sqrt{s}=10.58$ GeV.
We investigate its impact on the integrated cross sections,
and various types of distributions of $J/\psi$ at $B$ factory,
which is found to be modest.

Very recently the order-$v^2$ correction to $e^+e^-\to J/\psi gg$ at $B$ factory
has also been studied by He, Fan and Chao~\cite{He:2009uf}, who have
found similar magnitude of
the relativistic correction to the integrated cross section for unpolarized $J/\psi$.
Since none of the differential distributions for energy, angular and polarization of
$J/\psi$ have been explicitly given in \cite{He:2009uf},
it is not possible at this stage to make a detailed comparison between our results
and theirs. Nevertheless, these authors chose to expand $E_q$ around $m_c$
in powers of ${\mathbf q}^2$ in the amplitude,
which is opposite to the strategy employed in this work.
Most notably, it seems that relativistic correction effects associated with
the three-body phase space has been neglected in \cite{He:2009uf},
thus the exact agreement between our results and theirs
will not be expected.

\subsection{Choice of input parameters}

To make concrete predictions, we need specify various input parameters, in particular
the corresponding NRQCD production matrix elements.
The LO NRQCD matrix element $\langle {\mathcal O}_1^{J/\psi}
\rangle$, together with some specific combination of coupling
constants and mass scales will be frequently encountered in many
expressions for $J/\psi$ production rate. For notational compactness,
it is thus
convenient to lump them into a single factor:
\bqa
\check{\sigma}_0 & = &  { 256 \, \pi (e_c\alpha\alpha_s)^2 \over  27 \,
M_{J/\psi}\:s^2} \langle {\mathcal O}_1^{J/\psi} \rangle.
\label{def:combined:O:factor}
\eqa

We take $M_{J/\psi}=3.097$ GeV,
$\sqrt{s}=10.58$ GeV at $B$ factory energy, thus fix $r\equiv
M^2_{J/\psi}/s=0.0857$. For the NRQCD matrix elements, we quote the
values extracted from the recent Cornell-potential-model-based
analysis~\cite{Bodwin:2007fz}:
\begin{subequations}
\bqa
\langle {\mathcal O}_1^{J/\psi} \rangle &=&
0.440^{+0.067}_{-0.055}\;{\rm GeV}^3,
\label{O1:jpsi:num:value}
\\
\langle v^2 \rangle_{J/\psi} &=& 0.225^{+0.106}_{-0.088}.
\label{v2:jpsi:num:value}
\eqa
\end{subequations}
Note the uncertainties affiliated with each NRQCD matrix elements
are quite sizable.
With resort to the G-K relation (\ref{G-K:rel}),
one finds that
this value for $\langle v^2 \rangle_{J/\psi}$ corresponds to
the charm quark pole mass $m_c=1.39\pm 0.06$ GeV.

Targeting at a better accuracy, we also including the running effect
in the electromagnetic coupling, i.e., we take the fine structure
constant to be $\alpha(\sqrt{s})=1/130.9$, rather than the commonly
used $1/137$~\cite{Bodwin:2007ga}. For the strong coupling constant, we
take the central value $\alpha_s$ equal to $0.21$,
corresponding to choosing the renormalization scale $\mu$ at $\sqrt{s}/2$.
The corresponding uncertainty is estimated by varying this coupling
between $0.17$ to $0.26$, obtained by sliding the renormalization scale $\mu$
between $\sqrt{s}$ and $\sqrt{s}/4$~\cite{Bodwin:2007ga}.

With all these parameters specified, we find
\bqa
\check{\sigma}_0  &=& 0.150^{+0.115}_{-0.064} \;{\rm pb}.
\label{new:checksigma:value}
\eqa
The attached error comes from the uncertainties of $\alpha_s$
and of the NRQCD matrix element
$\langle {\mathcal O}_1^{J/\psi} \rangle$.

\subsection{The integrated production rate for $\bm{J/\psi}$ associated with light hadrons}
\label{int:unpol:cross:section:jpsi}

The lowest-order NRQCD predictions to the $J/\psi$ associated production rate have
been available for a long while~\cite{Keung:1980ev,Cho:1996cg,Yuan:1996ep,Baek:1998yf,Hagiwara:2004pf}.
For convenience of the reader, we collect in Appendix~\ref{appendix:all:cross:section} the expressions for the energy distribution of (un)polarized $J/\psi$, at LO as well as at NLO in $v^2$ .

One may attempt to directly integrate the differential $J/\psi$ spectrum over
the entire $J/\psi$ energy range to deduce the integrated $J/\psi$ cross section.
Unfortunately, it seems rather difficult to obtain the analytic expression by this way,
even for the LO cross section.
Fortunately, to this purpose, it is much more advantageous to
carry out the 3-body phase space integration in a route as specified in
(\ref{PS3-def-2-variables:altnative}), where the corresponding integration boundaries become simpler.
After some straightforward calculations, we find that
the LO integrated cross section for the unpolarized $J/\psi$ can be put in
the following compact form:
\bqa
\label{sig:int:unpol:v0}
& &  \sigma^{(0)} [J/\psi+X_{light}]
=  \check{\sigma}_0
\left\{{2-r-12 r^2+8r^3 \over 2(1-r)^2} \, {\rm arctanh}^2\sqrt{1-r}\right.
\\
&+& {4-9r+8r^2 \over (1-r)^{3/2}}\,{\rm arctanh}\sqrt{1-r}
+ {5-14r+3r^2\over 2(1-r)^2}\ln r
- {9(1-2r+2r^2)\over 2(1-r)} \bigg\}.
\nn
\eqa
We note that the analytic expression for $\sigma^{(0)}$ has already been
available recently~\cite{Gong:2009kp}. Our expression is in agreement with
equation~(2) in \cite{Gong:2009kp}, but appears to be simpler~\footnote{
This can be mainly attributed to the fact that a pair of dilogarithms
appearing in their formula can actually be transformed away, by exploiting
a sequence of identities about dilogarithms.}.

It is interesting to examine the asymptotic behavior of (\ref{sig:int:unpol:v0})
in the high energy limit $\sqrt{s}\gg M_{J/\psi}$:
\bqa
& &  \sigma^{(0)} [J/\psi+X_{light}]
= \check{\sigma}_0 \left[ {1\over 4}\ln^2 r + \left({1\over 2}-\ln 2\right)\ln r+
\ln^2 2 + 4\ln2-{9\over 2}+ O(r\ln^2r) \right].
\nn\\
\label{sig:int:unpol:v0:asym}
\eqa
Beside the power-law scaling contained in $\check{\sigma}_0$  ($\propto 1/s^2$),
the asymptotic behavior of the total cross section
is dominated by the double logarithm term.
This expression is superficially analogous to the NLO perturbative correction to the
exclusive double-charmonium production process $e^+e^-\to J/\psi+\eta_c$,
which also exhibits a double logarithm scaling~\cite{Gong:2007db}.

It is also of some interest to examine the opposite limit $r\to 1$,
in which the $J/\psi$ is produced just
above the kinematic threshold. The total cross section in this limit vanishes as
$\check{\sigma}_0 (1-r)/3 +  O((1-r)^2)$, which may reflect that gluon radiation
off the heavy quark is greatly damped in very restricted phase space.

Substituting $r=0.0857$ and the value of $\check{\sigma}_0$ given in
(\ref{new:checksigma:value}) into (\ref{sig:int:unpol:v0}), or equivalently,
numerically integrating the spectrum (\ref{dsig:dz:unpol:v0})
over the entire $J/\psi$ energy, we find the LO prediction to
the total cross section for $J/\psi$ associated with non-$c\bar{c}$ states
at $B$ factory is
\bqa
\sigma^{(0)}[J/\psi+X_{light}] &=& 0.200^{+0.153}_{-0.085} \;{\rm pb}.
\label{LO:int:Xectoin:unpol:jpsi}
\eqa
The error is solely due to the uncertainty in $\check{\sigma}_0$.

We now turn to the order-$v^2$ contribution to the
integrated cross section for producing the unpolarized $J/\psi$.
Unlike in the LO case, the corresponding analytical expression
is too complicated to be presented here, and we are content with providing
numerical result only.
Taking the value of $\check{\sigma}_0$ from (\ref{new:checksigma:value}),
together with the ratio of the NRQCD matrix elements
$\langle v^2 \rangle_{J/\psi}$ in (\ref{v2:jpsi:num:value}),
integrating the order-$v^2$ correction to the spectrum (\ref{dsig:dz:unpol:v2})
over the entire $J/\psi$ energy range, we find
\bqa
\sigma^{(2)}[J/\psi+X_{light}] &=& 0.061^{+0.097}_{-0.040} \;{\rm pb},
\label{NLO:int:Xectoin:unpol:jpsi}
\eqa
where the attached error is due to the uncertainties in $\check{\sigma}_0$
and in $\langle v^2 \rangle_{J/\psi}$. It is clear to see that
for the central value of the predictions, the inclusion of the order-$v^2$ correction
enhances the LO result by about 30\%.
This seems in conformity to the naive expectation about the size of relativistic correction
for charmonium system.
It is interesting to note that, the central value of the relative order-$v^2$ contribution
seems even slightly larger than the recently-computed NLO perturbative correction,
which enhances the LO result by about 20\%~\cite{Ma:2008gq,Gong:2009kp}.
But fairly speaking, the effects of both types of corrections are not significant.

The sum of (\ref{LO:int:Xectoin:unpol:jpsi}) and
(\ref{NLO:int:Xectoin:unpol:jpsi}) turns to be
\bqa
(\sigma^{(0)} + \sigma^{(2)})[J/\psi+X_{light}] &=& 0.261^{+0.250}_{-0.125} \;{\rm pb}.
\label{LO+NLO:int:Xectoin:unpol:jpsi}
\eqa
Compared with the latest \textsc{Belle} measurements for prompt $J/\psi$ production rate associated
with light hadrons, (\ref{belle:latest:jpsi:plus:light}), we find rough agreement between (\ref{LO+NLO:int:Xectoin:unpol:jpsi}) and the data with large uncertainty.
This agreement could be even more satisfactory
if further including the NLO perturbative correction
and the feeddown contribution from higher charmonium states.

In light of this rough agreement achieved by the color-singlet contribution alone,
one important question is to ask how much room is left for the color-octet contribution to
inclusive $J/\psi$ production at $B$ factory.
It seems fair to state that earlier estimates of
its contribution~\cite{Braaten:1995ez,Yuan:1997sn,Fleming:2003gt}
may turn out to be overly optimistic.
Nevertheless, we would like to caution that, our predictions, both the LO one in
(\ref{LO:int:Xectoin:unpol:jpsi}), and the NLO one in (\ref{NLO:int:Xectoin:unpol:jpsi}),
are subject to large theoretical uncertainty, so we are unable to
draw any firm conclusion about the actual size of the color-octet contribution.

\subsection{Energy spectrum of unpolarized $\bm{J/\psi}$}

\begin{figure}[htb]
\centerline{
\includegraphics[height=4.9cm]{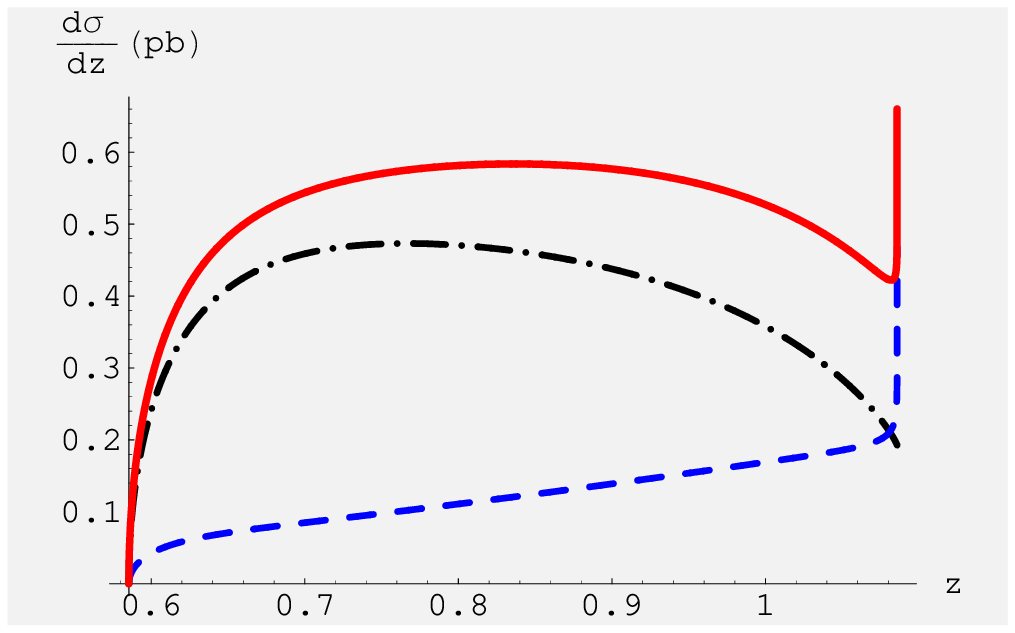}
\quad\quad
\includegraphics[height=4.9cm]{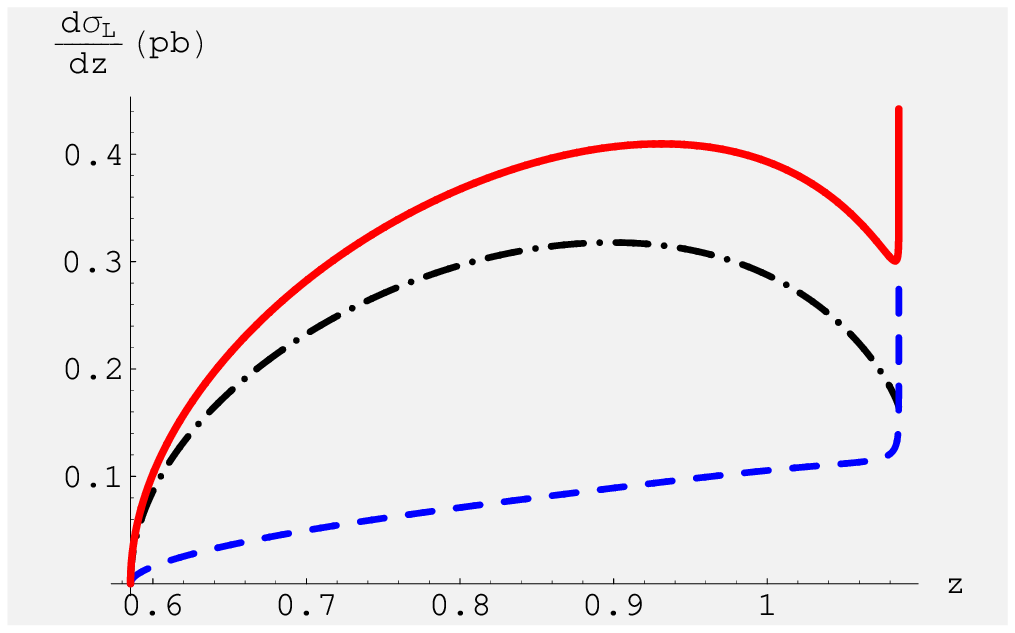}}
\caption{The energy spectra of the unpolarized $J/\psi$ (left panel) and
longitudinally polarized $J/\psi$ (right panel) associated with light hadrons
at the energy of $B$ factory. The dot-dashed curve represents
$d\sigma^{(0)}/dz$, the dashed curve represents $d\sigma^{(2)}/dz$,
and the solid curve represents their sum.
For simplicity, in all the figures in this work,
we have taken only the central values of the input parameters
and not drawn the error band.
\label{dsig:dz:jpsi:v0:v2}}
\end{figure}

Aside from the total production rate of $J/\psi$, it is also useful to look closely into
the differential observable. As a matter of fact, $B$ factory experiments have already
measured various types of $J/\psi$ distributions.
In this subsection we investigate the effect of first-order relativistic correction for the
energy distribution of unpolarized $J/\psi$.
In Figure~\ref{dsig:dz:jpsi:v0:v2}, we display the energy spectrum of
unpolarized $J/\psi$ at $B$-factory energy, including both the LO result and
the first-order relativistic correction.

As one can tell from Figure~\ref{dsig:dz:jpsi:v0:v2},
the LO distribution admits a finite limit when the $J/\psi$ energy
approaches its maximum:
\bqa
\left.{d \sigma^{(0)}\over dz} \big[e^+e^-\to J/\psi+gg
\big]\right|_{z\to 1+r} \longrightarrow  \check{\sigma}_0 {1+2r\over 1-r}.
\label{dsig0:dz:upper:end}
\eqa

A novel feature of relative order-$v^2$ contribution is that, as can be clearly
seen in Figure~\ref{dsig:dz:jpsi:v0:v2}, the spectrum has a sharp rise
near the very upper end of the $J/\psi$ spectrum.
After some straightforward manipulation on
the analytic expression of order-$v^2$ contribution,
which is recorded in equation (\ref{dsig:dz:unpol:v2}),
we find the following limiting value near the endpoint:
\bqa
\left.{d \sigma^{(2)}\over dz} \big[e^+e^-\to J/\psi+gg
\big]\right|_{z\to 1+r} &\longrightarrow& {\check{\sigma}_0 \, \langle v^2 \rangle_{J/\psi}\over 3}
\nn\\
&\times& {9-23 r - 10 r^2 -12 r (1+r) \ln\big[{r\,(1+r-z)\over
(1-r)^2}\big] \over (1-r)^2}.
\label{end:point:unpol:v2}
\eqa
Clearly the endpoint singularity is of the form $\ln(1+r-z)$.

The logarithmic divergence near the endpoint is not something new.
It simply signals the breakdown of the NRQCD expansion near the kinematic boundary, as a result
we should no longer trust our prediction in this region.
Recall that for the NLO perturbative correction to the same process,
the logarithmic singularity of $\ln(1+r-z)$ is also expected to
appear near the maximum of $J/\psi$ energy~\cite{Lin:2004eu,Leibovich:2007vr}.
However, it is worth mentioning that, the $\ln(1+r-z)$ has rather different origin
for both types of corrections. For the NLO perturbative correction, the $\ln(1+r-z)$ term
should be attributed to the collinear singularity associated with
the gluonic jet recoiling against $J/\psi$.
The reason is that, at LO in $v$, the soft gluon cannot resolve the
color-singlet $c\bar{c}$ pair (color-transparency), as a result
the net contributions from soft gluons cancel,
so the logarithm can be only of the collinear origin~\footnote{It seems enlightening
to contrast the single collinear logarithm
associated with the color-singlet channel at LO in $v$ with
the Sudakov double logarithm associated with the color-octet channel~\cite{Fleming:2003gt}.}.
However, for the contribution from relativistic correction, this endpoint
singularity comes from the region where one of the recoiling gluon becomes soft.
Since we have gone beyond the LO in $v$, the color-singlet $c\bar{c}$ pair
could still develop a nonzero color dipole,
therefore it may strongly interact with the soft gluons.
Therefore it is natural to identify this resulting $\ln(1+r-z)$ with the soft origin.
It is interesting to ask whether the method presented in \cite{Lin:2004eu,Leibovich:2007vr},
which combine NRQCD and the soft-collinear effective theory,
can be generalized to resum those types of logarithm in (\ref{end:point:unpol:v2}) to
all orders in $\alpha_s$, to render the $J/\psi$ energy spectrum well-behaved
near the end point region.

Note this endpoint singularity is integrable, therefore we are still able to
obtain a finite order-$v^2$ correction to the integrated cross section
(see (\ref{NLO:int:Xectoin:unpol:jpsi})).
This is similar to quarkonium semi-inclusive radiative decay $J/\psi\to \gamma+X$, where
the order-$v^2$ correction to the photon spectrum also develops
an integrable endpoint singularity.
Nevertheless in that case, at relative order $v^4$, the photon spectrum would develop a linear infrared
divergence near the end point, which results in a
logarithmic divergence for the integrated decay rate~\cite{Bodwin:2002hg}.
It is the color-octet mechanism that should be invoked to tame this infrared divergence.
In our case, we expect the exactly same pattern will occur. That is,
at $O(v^4)$, the $J/\psi$ energy spectrum would develop a linear endpoint singularity,
consequently the integrated cross section would contain a logarithmic infrared divergence,
which must in turn be cured by including the color-octet contribution.

\subsection{Polarization distribution of $\bm{J/\psi}$}


\textsc{Babar} and \textsc{belle} collaborations can also determine the
polarization of $J/\psi$ as a function of its energy by measuring the muons'
angular distribution from $J/\psi\to \mu^+\mu^-$. The commonly
used polarization parameter is defined by
\bqa
\alpha(z) & = & {d\sigma/dz-3 d\sigma_L/dz \over
d\sigma/dz+d\sigma_L/dz},
\label{alpha:definition}
\eqa
where $d\sigma_L/dz $ signifies the differential cross section for
producing a longitudinally-polarized $J/\psi$. $\alpha=1$ and $-1$
correspond to 100\% transversely- and longitudinally-polarized, whereas
$\alpha=0$ corresponds to 100\% unpolarized.

To deduce the function $\alpha(z)$, it is necessary to know the expression
for $d\sigma_L/dz $. The analytical expressions for this distribution,
at both LO and NLO in $v^2$, have been given in
Appendix~\ref{appendix:all:cross:section}. Moreover, both the LO and NLO
contributions to the energy spectrum for the longitudinally-polarized $J/\psi$
at $B$-factory is shown in Figure~\ref{dsig:dz:jpsi:v0:v2}.

Let us first investigate the integrated cross section for producing a
longitudinally polarized $J/\psi$.
As in the unpolarized case discussed in Section~\ref{int:unpol:cross:section:jpsi},
if one carries out the 3-body phase-space integration following the order specified in
(\ref{PS3-def-2-variables:altnative}), the LO integrated cross section for
the longitudinally-polarized $J/\psi$ can also be put in a closed form:
\bqa
& &  \sigma^{(0)}_L [J/\psi(\lambda=0)+X_{light}]
= \check{\sigma}_0 \left\{ {4-2r-4 r^2-3 r^3+3r^4 \over 4(1-r)^2 }\, {\rm arctanh}^2\sqrt{1-r} \right.
\nn \\
&-& { 4-4 r + r^2 - 3r^3 \over 2(1-r)^{3/2} }\, {\rm arctanh}\sqrt{1-r}
+ {2-10 r + 7 r^2- 6 r^3 + 3r^4 \over 4(1-r)^2}\,\ln r
\nn \\
&+& {\sqrt{r}\,(6- r + 3 r^2) \over 4}\, \bigg(\, {\rm Li}_2(\sqrt{r})-
 {\rm Li}_2(-\sqrt{r}) - {\rm arctanh} \sqrt{r}\,\ln r
\bigg )
\nn \\
&+&\left.
{\pi^2\over 16}\, \left(2-12\sqrt{r}+3r+2r^{3 \over 2}+3r^2-6r^{5 \over 2}\right)+
{8-19r+11r^2-6r^3 \over 4(1-r)}\right\},
\label{sig:int:long:pol:v0}
\eqa
where ${\rm Li}_2$ stands for the dilogarithm.
This analytic expression has not been known previously.
A nontrivial check of the correctness of this result is to examine its
threshold behavior. In the limit $r\to 1$,  $\sigma^{(0)}_L$  approaches zero
as $\check{\sigma}_0 {1-r \over 9} + O((1-r)^2)$, as expected,
one-third of that for the polarization-summed case.

It is of interest to ascertain the asymptotic behavior of $\sigma_L^{(0)}$ in
the high energy limit $\sqrt{s}\gg M_{J/\psi}$:
\bqa
& & \sigma^{(0)}_L [J/\psi(\lambda=0) +X_{light}]
\nn\\
&=&\check{\sigma}_0 \bigg[ {1\over 4}\ln^2 r + \left({3\over 2}-\ln 2\right)\ln r
+\ln^2 2 -2\ln2+ 2+{\pi^2\over 8}+ O(\sqrt{r}) \bigg].
\label{sig:int:long:jpsi:v0:asym}
\eqa
Note the leading double logarithm appearing here has the same coefficient as in
the polarization-summed case, (\ref{sig:int:unpol:v0:asym})~\footnote{In contrast to (\ref{sig:int:unpol:v0:asym}),
the leading correction to this asymptotic expression is of relative order $1/\sqrt{s}$,
instead of $1/s$.}. One can then readily infer the asymptotic behavior of the
transversely-polarized $J/\psi$ production rate:
$\sigma_T^{(0)}\equiv \sigma^{(0)}-\sigma_L^{(0)} \to
\check{\sigma}_0\left[\ln{1\over r}+6\ln2-{13\over 2}-{\pi^2\over 8}\right]$,
only exhibiting a single-logarithm scaling.

Substituting $r=0.0857$ into (\ref{sig:int:long:pol:v0}) and using the
value of $\check{\sigma}_0$ given in
(\ref{new:checksigma:value}), or straightforwardly
integrating the spectrum (\ref{dsig:dz:long:pol:v0})
over the entire $J/\psi$ energy numerically,
we find the LO prediction to the integrated rate for producing longitudinally-polarized
$J/\psi$ in association with non-$c\bar{c}$ states at $B$ factory to be
\bqa
\sigma_L^{(0)}[J/\psi+X_{light}] &=& 0.128^{+0.098}_{-0.054} \;{\rm pb}.
\label{LO:int:Xectoin:longpol:jpsi}
\eqa
The error originates solely from the uncertainty in $\check{\sigma}_0$.
According to (\ref{alpha:definition}), and using the central value of
(\ref{LO:int:Xectoin:unpol:jpsi}) and (\ref{LO:int:Xectoin:longpol:jpsi}),
we find the $\alpha=-0.56$ averaged over the entire $J/\psi$ energy range.

For the order-$v^2$ contribution to $\sigma_L$,
the corresponding analytic expression is too involved, if not impossible,
to deduce, so we are content with providing numerical result only.
Using $\langle v^2 \rangle_{J/\psi}$ as given in (\ref{v2:jpsi:num:value}),
integrating the order-$v^2$ correction (\ref{dsig:dz:long:pol:v2})
over the entire $J/\psi$ energy range, we get
\bqa
\sigma_L^{(2)}[J/\psi+X_{light}] &=& 0.037^{+0.059}_{-0.024} \;{\rm pb}.
\label{v2:int:Xectoin:longpol:jpsi}
\eqa
The attached error comes from the uncertainties in $\check{\sigma}_0$
and $\langle v^2 \rangle_{J/\psi}$.
For the central values of the predictions,
inclusion of the order-$v^2$ correction
enhances the LO cross section by about 29\%, which has a
very similar magnitude of enhancement as for the unpolarized $J/\psi$.
This is again in accordance with the naive expectation about
the size of relativistic correction for charmonium system.

The sum of (\ref{LO:int:Xectoin:longpol:jpsi}) and
(\ref{v2:int:Xectoin:longpol:jpsi}) is
\bqa
(\sigma_L^{(0)} + \sigma_L^{(2)})[J/\psi+X_{light}] &=&
0.165^{+0.157}_{-0.079} \;{\rm pb}.
\label{sum:int:Xectoin:longpol:jpsi}
\eqa
Including the order-$v^2$ correction, the central value of the
average polarization variable $\alpha$ shifts from $-0.56$ to $-0.55$.
Hence the relativistic correction has a rather minor effect
in changing the polarization of $J/\psi$.

Now let us examine the differential distribution $d \sigma^{(0)}_L/dz$.
As can be seen from Figure~\ref{dsig:dz:jpsi:v0:v2},
or can be directly inferred from (\ref{dsig:dz:long:pol:v0}),
the LO distribution has a finite limit when the energy of the
longitudinally-polarized $J/\psi$ approaches its maximum:
\bqa
\left.{d \sigma^{(0)}_L \over dz} \big[e^+e^-\to
J/\psi(\lambda=0)+gg \big]\right|_{z\to 1+r} \longrightarrow
\check{\sigma}_0 {1\over 1-r}.
\label{dsiglong0:dz:upper:end}
\eqa
From (\ref{dsig0:dz:upper:end}) and (\ref{dsiglong0:dz:upper:end}),
it is ready to see that, at the endpoint $z=1+r$, the LO polarization variable,
$\alpha^{(0)}$, approaches the constant $-{1-r \over 1+r}$.

As can be seen in Fig.~\ref{dsig:dz:jpsi:v0:v2}, the
order-$v^2$ correction to the energy spectrum of the
longitudinally polarized $J/\psi$ also diverges logarithmically near the
upper end. After some manipulation on equation (\ref{dsig:dz:long:pol:v2}),
we find the following limiting behavior:
\bqa
\left.{d \sigma^{(2)}_L \over dz} \big[e^+e^-\to
J/\psi(\lambda=0)+gg \big]\right|_{z\to 1+r} & \longrightarrow & {
\check{\sigma}_0 \, \langle v^2
\rangle_{J/\psi}\over 3}
\nn\\
&\times& {5-17r+4 r^2 -8r \ln\big[{r\,(1+r-z)\over (1-r)^2}\big]
\over (1-r)^2}.
\eqa
The situation very much resembles that for the unpolarized $J/\psi$.
One can refer to the paragraphs after (\ref{end:point:unpol:v2}) for similar discussions.

\begin{figure}[htb]
\centerline{
\includegraphics[height=6 cm]{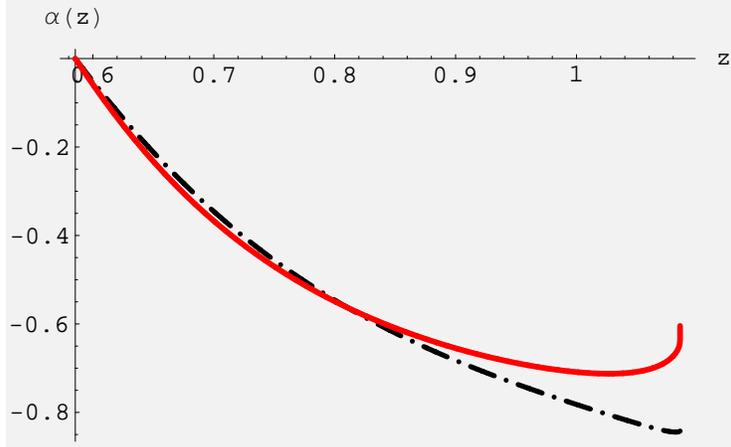}
} \caption{Profile of the $J/\psi$ polarization parameter $\alpha(z)$ in associated production
with light hadrons at $\sqrt{s}=10.58$ GeV. The dot-dashed curve represents the leading
order prediction $\alpha^{(0)}$, whereas the solid curve represents the corresponding one
including the $O(v^2)$ effect.
\label{alpha:v0:vs:v2} }
\end{figure}

In Figure~\ref{alpha:v0:vs:v2}, we also display how the polarization
parameter $\alpha$ varies with the $J/\psi$ energy. Clearly, the
inclusion of relativistic correction seems to have a minor impact in most of
the region of $z$, except increasing it modestly near the upper end.

\subsection{Angular-Energy distribution of $\bm{J/\psi}$}

Experimentally it is also possible to measure the production rate for $J/\psi$
in $e^+e^-$ annihilation that is differential in $\cos\theta$, the cosine of the
angle between the momentum of $J/\psi$ and the incident $e^-$ beam in the
laboratory frame. It is thus theoretically interesting to study the
differential angular distribution of $J/\psi$.
As pointed out in (\ref{double:diff:distr:form}),
for inclusive $J/\psi$ production in $e^+e^-$ annihilation,
general consideration constrains the double differential distribution
of the following form~\cite{Cho:1996cg}:
\bqa
{d \sigma^{(i)} \over dz d \cos\theta} \big[e^+e^-\to J/\psi+X\big]
&=& S^{(i)}(z)\big[ 1+A^{(i)}(z)\cos^2\theta \big],
\label{double:diff:distribution:v0:v2}
\eqa
where $A(z)$ is a angular parameter that satisfies $|A(z)|\le 1$.

The analytic expressions at LO in $v$, $S^{(0)}(z)$ and $A^{(0)}(z)$  have been
known long ago. The closed forms of the order-$v^2$ contributions,
$S^{(2)}(z)$ and $A^{(2)}(z)$, are derived in this work for the first time.
For completeness, we reproduce all of them
in Appendix~\ref{appendix:all:cross:section}.
From Eqs.~(\ref{dsig:dz:dcosth:S:v0}) and (\ref{dsig:dz:dcosth:SA:v0}),
one finds that the LO double differential spectrum
admits a finite limit near the upper end~\cite{Braaten:1995ez}:
\bqa
\left.{d \sigma^{(0)}\over dz\, d \cos\theta} \big[e^+e^-\to
J/\psi+gg \big]\right|_{z\to 1+r}  &\longrightarrow &
{3\,\check{\sigma}_0 \over 4}
\,\left({1+r\over 1-r}- \cos^2\theta\right),
\eqa
which implies that at the endpoint $z=1+r$, $A^{(0)}=-{1-r \over 1+r}$.

\begin{figure}[htb]
\centerline{
\includegraphics[height=6 cm]{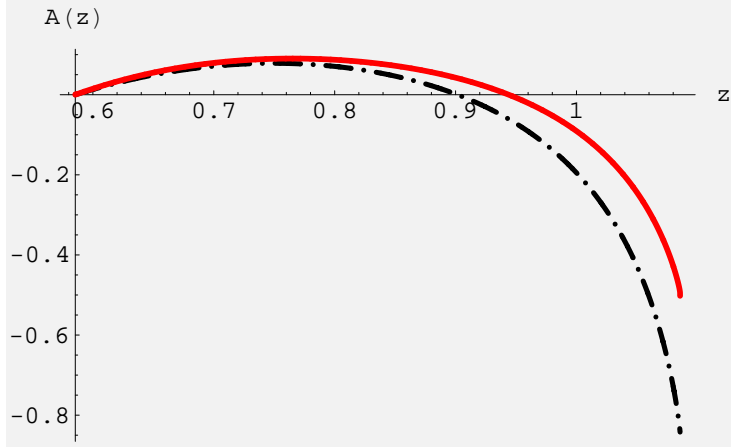}
} \caption{Profile of the $J/\psi$ angular distribution parameter $A(z)$
in associated production with light hadrons at $\sqrt{s}=10.58$ GeV.
The dot-dashed curve represents the leading-order prediction $A^{(0)}$,
whereas the solid curve represents $A^{v^2}(z)$
defined in (\ref{A:corr:NLO:v2}), which has included the order-$v^2$
effect.
\label{A:v0:vs:v2} }
\end{figure}

In Figure~\ref{A:v0:vs:v2}, we display the angular function $A(z)$ at energy
of the $B$ factory, $\sqrt{s}=10.58$ GeV. Both the LO prediction and
that including the first-order relativistic correction are shown.
Note that the correct $A(z)$ incorporating the order-$v^2$
effect is given by~\cite{Cho:1996cg}
\bqa
A^{v^2}(z) &=& {S^{(0)}(z) A^{(0)}(z) + S^{(2)}(z) A^{(2)}(z)
\over S^{(0)}(z)+S^{(2)}(z)}.
\label{A:corr:NLO:v2}%
\eqa
As can be seen in Figure~\ref{A:v0:vs:v2}, including the
first-order relativistic correction seems to have modest effect, which
only slightly softens the angular distribution
near the upper end.

Next let us inspect the end-point behavior of the functions
$S^{(2)}(z)$ and $A^{(2)}(z)$.
After some straightforward algebra from Eqs.~(\ref{dsig:dz:dcosth:S:v2}) and
(\ref{dsig:dz:dcosth:SA:v2}), we find the following limiting behaviors when
$z$ approaches its maximum:
\begin{subequations}
\bqa
S^{(2)}(z) & \longrightarrow & {\check{\sigma}_0 \, \langle v^2 \rangle_{J/\psi} \over 4}
\,{5-16r-5 r^2 -2r(5+3r) \ln\big[{r\,(1+r-z)\over (1-r)^2}\big]\over
(1-r)^2},
\\
S^{(2)}(z) A^{(2)}(z)& \longrightarrow & {\check{\sigma}_0 \, \langle v^2 \rangle_{J/\psi}\over 4
}\, {3+5 r + 6r \ln\big[{r\,(1+r-z)\over (1-r)^2}\big] \over 1-r}.
\eqa
\end{subequations}
These emerging logarithmic divergences near the endpoint
simply reflects that the differential cross section diverges
in that region. However, according to Eq.~(\ref{A:corr:NLO:v2}), the
angular distribution $A^{v^2}(z)$, defined as a ratio,
still remains a finite and smooth function near the very upper end.

It is worth mentioning that \textsc{Belle} collaboration has recently measured the average
angular variable for the $J/\psi$ production in association with noncharmful states,
$\overline{A}_{\rm exp}= 5.2^{+6.1}_{-2.4}(0.3)$~\cite{Pakhlov:2009nj}.
Theoretically, it is straightforward to define the corresponding
$\overline{A}$ by integrating
(\ref{double:diff:distribution:v0:v2}) over $z$:
\bqa
{d \sigma^{(i)} \over d \cos\theta} \big[e^+e^-\to J/\psi+X_{light}\big]
&=& \overline{S}^{(i)}\big[ 1+ \overline{A}^{(i)}\cos^2\theta \big],
\eqa
From Eqs.~(\ref{dsig:dz:dcosth:S:v0}) and (\ref{dsig:dz:dcosth:SA:v0}), and inserting
$r=0.0857$, we find the LO NRQCD prediction is $\overline{A}^{(0)}= -0.037$.
Notwithstanding the large experimental uncertainty, this prediction is in apparent disagreement
with the \textsc{Belle} measurement, even the sign is opposite.
Subsequent studies reveal that including the NLO perturbative correction does not help to resolve this discrepancy~\cite{Gong:2009ng}.

One may naturally wonder whether implementing the relativistic correction will bring
the NRQCD prediction closer to the data or not. In analogy with (\ref{A:corr:NLO:v2}),
we introduce a new average angular variable that incorporates the  $O(v^2)$  effect:
\bqa
\overline{A}^{\,v^2}  &=& {\overline{S}^{(0)}  \overline{A}^{(0)}  +
\overline{S}^{(2)} \overline{A}^{(2)}
\over \overline{S}^{(0)} + \overline{S}^{(2)}}.
\label{A:aver:corr:NLO:v2}%
\eqa
Starting from Eqs.~(\ref{dsig:dz:dcosth:S:v2}) and
(\ref{dsig:dz:dcosth:SA:v2}), and adopting
the central value of $\langle v^2 \rangle_{J/\psi}$ tabulated in
(\ref{v2:jpsi:num:value}), we then find $\overline{A}^{\,v^2}=0.0011$,
which now has the same sign as the measured value, though still differs
considerably in the absolute magnitude.
Therefore, including NLO relativistic correction (plus perturbative correction)
seems not to be sufficient to explain the data.
It remains to be a challenge how to correctly account for the measured
angular distribution in the $J/\psi+X_{light}$ channel
within the NRQCD factorization framework.

\section{Discussion and summary}
\label{summary:outlook}

In this work, we have introduced a somewhat heterodox
NRQCD matching strategy, which is particularly suitable for calculating the
relativistic correction to
(inclusive) quarkonium production and decay processes
with involved kinematics in the color-singlet channel.
The great advantage of our approach over
the orthodox matching strategy is that, it can take into account the
relativistic correction effect in the phase space integration
with much ease, thanks to the Gremm-Kapustin relation.
As a nontrivial application of this method,
we have systematically investigated the relative order-$v^2$ correction
to the inclusive $J/\psi$ production associated with light hadrons at $B$ factories.
We have found that it can modestly enhance the lowest-order NRQCD prediction for the
integrated $J/\psi$ cross section, about 30\% if we choose the relativistic
correction matrix element as specified in \cite{Bodwin:2007fz}.
We find its impact on the $J/\psi$ polarization and angular distributions is quite minor.
The magnitude of the order-$v^2$ correction seems to be comparable with
that of the respective NLO perturbative correction.
We would like to caution that, our predictions of the order-$v^2$ correction
are likely subject to large theoretical uncertainty.
In particular, some intrinsic uncertainty related to the relative
order-$v^2$ NRQCD matrix element seems to restrict our ability to make precise predictions
for the relativistic correction.
Since the corrections computed in this work has only a modest effect,
we feel unable to draw any sharp conclusion,
especially for the actual size of the color-octet contribution.

Of the special theoretical interest, is the logarithmic divergence near the upper end
of the $J/\psi$ spectrum found in this work for the order-$v^2$ contribution.
It is desirable to extend the
theoretical framework developed in \cite{Fleming:2003gt,Lin:2004eu,Leibovich:2007vr}
beyond the LO in $v$,
to see whether such type of soft endpoint logarithms
can be resummed to all orders in $\alpha_s$, to render the
$J/\psi$ energy spectrum well-behaved in the end point region.

Another interesting direction is to incorporate the order-$v^4$ correction
to the process considered in this work. We expect that the perturbative matching approach
described in this work, after some straightforward extension,
is well-suited to achieve this goal.
At $O(v^4)$,  the $J/\psi$ energy spectrum is expected to exhibit
a linear divergence near the upper end point, and consequently,
the integrated cross section will be logarithmically divergent.
It will be interesting to see how the color-octet contribution from the
${}^3P_J^{(8)}$ NRQCD production operator is explicitly put into work
to tame this infrared divergence.

\begin{acknowledgments}
First I wish to thank Jian-Xiong Wang for his inquiry in spring of 2008
that stimulated me to initiate this research, and for many
informative exchanges concerning $J/\psi$ production in various collision
experiments.
It is also a pleasure to acknowledge Bin Gong, Adam Leibovich,
Jian-Wei Qiu and Guo-Huai Zhu for
valuable communications on related topics.
I would also like to take this opportunity to thank KITPC at Beijing
for hosting an enjoyable program entitled
{\it Effective Field Theories in Particle and Nuclear physics}
(Aug.~3--Sep.~11, 2009), during which part of this manuscript was written.
This research was supported in part by the National Natural Science
Foundation of China under grants No.~10875130, 10935012,
and by the Project of Knowledge Innovation Program (PKIP) of
Chinese Academy of Sciences, Grant No. KJCX2.YW.W10.

\end{acknowledgments}

\appendix

\section{Miscellaneous formulas for inclusive
$\bm{J}\bm{/}\bm{\psi}$ production associated with light hadrons in
$\bm{e^+ e^-}$ annihilation
\label{appendix:all:cross:section}%
}

In this section, we collect the analytic expressions for
various types of distributions for $J/\psi$ associated
production with light hadrons
in $e^+e^-$ annihilation. Each type of differential cross section
is understood to contain two parts: $d\sigma = d\sigma^{(0)}+d\sigma^{(2)}$, which represent
the leading order contribution, and the contribution of relative
order-$v^2$, respectively. We emphasize that it is the physical
$J/\psi$ mass, rather than the charm quark mass, that enters into
the formulas of each part.

\subsection{Energy distribution of unpolarized $J/\psi$
\label{}%
}

The energy spectrum of unpolarized $J/\psi$ at LO in $v$ reads:
\bqa
& & {d \sigma^{(0)}\over dz} \big[e^+e^-\to J/\psi+gg \big]
\nn \\
& =&  \check{\sigma}_0 {1\over
(2-z)^2 (z -2r)^3}
\nn\\
&\times& \left\{ (z-2r)\sqrt{z^2-4 r}\,\left[ \, 4(1+5 r+7r^2+4 r^3)
\right.\right.
\nn \\
&-& \left. 12(1+r)(1+2r) z +(13+14r)z^2 -4z^3\right]
\nn\\
&+& 4(1+r-z) \left[ 2r(1-r)(1+8r+4 r^2)- 2r(5-2r-6 r^2) z \right.
\nn\\
&+& \left.\left. (1+r-5r^2)z^2 \right]\,
\ln\left({z-2r+\sqrt{z^2-4r}\over z-2r-\sqrt{z^2-4r}}\right)
\right\},
\label{dsig:dz:unpol:v0}
\eqa
where the quantity $\check{\sigma}_0$ has
been defined in (\ref{def:combined:O:factor}). This expression agrees
with the result given in \cite{Keung:1980ev}, but differs from
Ref.~\cite{Cho:1996cg,Yuan:1996ep,Baek:1998yf} by an overall
constant.

The first-order relativistic correction to the energy spectrum of
unpolarized $J/\psi$ reads:
\bqa
& & {d \sigma^{(2)}\over dz} \big[e^+e^-\to J/\psi+gg \big]
\nn\\
&= & {\check{\sigma}_0 \,
\langle v^2 \rangle_{J/\psi}\over 3} {1\over (2-z)^4 (z -2 r)^5}
\nn \\
& \times &  \left\{
 (z-2 r)\sqrt{z^2-4 r}
\left[\, 64 r(3+11 r-2r^2-4 r^3-20 r^4-15 r^5) \right.\right.
\nn \\
&-& 32r (22+30 r- 21 r^2 - 91 r^3- 89 r^4-7 r^5) z
\nn \\
&-& 16 (1-48r-9 r^2+ 171 r^3 +213 r^4 +35 r^5) z^2
\nn \\
&+&16 (4-22r+66 r^2+133 r^3+ 35r^4) z^3
\nn \\
&-&  4 (18+15r+170 r^2+ 70 r^3) z^4
\nn\\
&+& \left. 4 (11+15 r+ 16 r^2) z^5 -(11-2r)z^6 \,\right]
\nn \\
&+& 4 \, \left[ \, 32 r^2 (1+r)(3+9r-6r^2+9r^3+14 r^4+15r^5)\right.
\nn \\
&-& 16r^2 (24+50r +21 r^2+ 124 r^3+ 228 r^4+ 126 r^5+7r^6) z
\nn \\
&-& 8r (3-63 r-103 r^2-278 r^3- 697 r^4-469 r^5-49r^6) z^2
\nn \\
&+& 8r (9-50 r-186 r^2-549 r^3 -477 r^4- 77 r^5) z^3
\nn \\
&+& 2 (2-37 r+248 r^2+993 r^3+1122 r^4 +272 r^5) z^4
\nn\\
&-& 4(2-2r+124r^2+192 r^3+71 r^4) z^5
\nn \\
&+& (7+26r+140r^2+87 r^3)z^6
\nn\\
 & - & \left.\left. (3+2r+14 r^2)z^7 \,\right]\,
\ln\left({z-2r+\sqrt{z^2-4r}\over z-2r-\sqrt{z^2-4r}} \right)
\right\},
\label{dsig:dz:unpol:v2}
\eqa
where $\langle v^2 \rangle_{J/\psi}$ has been introduced in
(\ref{v2:jpsi:definition}).

\subsection{Energy distribution of longitudinally-polarized $J/\psi$
\label{}%
}

The energy spectrum of the longitudinally-polarized $J/\psi$ at LO in
$v$ reads:
\bqa
& & {d \sigma^{(0)}_L \over dz} \big[e^+e^-\to J/\psi(\lambda=0)+gg
\big]
\nn\\
 &=&\check{\sigma}_0
{1\over (2-z)^2 (z -2r)^3 (z^2-4r)} \left\{ (z-2r)\sqrt{z^2-4r}
\right.
\nn\\
&\times & \left[ -8r^2(9+9r+r^2+3r^3)+ 16r (2+8 r+ r^2 +3 r^3) z
\right.
\nn \\
&-& \left. 4(1+10r+3r^2+8r^3) z^2 +4(1-r+2r^2)z^3+z^4 \right]
\nn \\
&+& 4 (1+r-z) \left[-4r^3(1-r) (9+2r+3r^2)+ 8 r^2(2-4r-3r^3)z
\right.
\nn \\
&-& 2r(1-16 r -2r^2 -9 r^3) z^2-2r (5+3 r +3r^2) z^3
\nn \\
&+& \left.\left. (1+r+r^2)
z^4\right]\,\ln\left({z-2r+\sqrt{z^2-4r}\over z-2
r-\sqrt{z^2-4r}}\right) \right\}.
\label{dsig:dz:long:pol:v0}
\eqa

The order-$v^2$ correction to the energy spectrum of
the longitudinally-polarized $J/\psi$ is
\bqa
& & {d \sigma^{(2)}_L \over dz}[e^+e^-\to J/\psi(\lambda=0) + gg]
\nn\\
&=& {\check{\sigma}_0 \,
\langle v^2
\rangle_{J/\psi}\over 3}\, %
{1\over (2-z)^4 (z -2 r)^5(z^2-4r)}
\nn \\
&\times &  \left\{(z-2 r)\sqrt{z^2-4 r}\left[ -128 r^3 (15+4r+5 r^2+
r^3-4 r^4-15 r^5) \right.\right.
\nn \\
&+& 64 r^2 (6+62r+30 r^2+39 r^3 + 21 r^4-101 r^5-9r^6) z
\nn \\
&- & 32 r^2 (6+122r+153 r^2+215 r^3-247 r^4-57 r^5) z^2
\nn \\
&-& 16 r (2+6 r-275 r^2 - 571 r^3+ 233 r^4+141 r^5) z^3
\nn \\
&-& 16 (1+7 r+ 51 r^2 + 358 r^3+ 16 r^4 - 83 r^5) z^4
\nn \\
& + & 8 (6+45r + 205 r^2 + 113 r^3 -41 r^4) z^5
\nn \\
&-& 4(14 + 91 r+ 76r^2 + 3 r^3)z^6
\nn \\
& + & \left. 4 (8+ 15 r + 5 r^2) z^7 - (3+4r) z^8 \right]
\nn\\
&+& 4 \left[ -64 r^4 (15 + 9 r + 3 r^2 -2 r^3 - r^4 + 9 r^5 + 15
r^6) \right.
\nn \\
&+& 32 r^3 (6 + 74 r + 32 r^2 - 5 r^3 -28 r^4+ 26 r^5+134 r^6 + 9
r^7) z
\nn \\
&-& 16 r^4 (108 - 9 r- 166 r^2-118 r^3 + 478  r^4 + 75 r^5) z^2
\nn \\
&-& 8 r^2 (6+116 r + 71 r^2 + 492 r^3 + 784 r^4- 840 r^5- 261 r^6)
z^3
\nn \\
&-& 8 r (1-21 r-195 r^2- 437 r^3 -947 r^4 + 308 r^5 + 243 r^6) z^4
\nn\\
&-& 4r(2+63 r+ 497 r^2 +1253 r^3 + 133 r^4 - 252 r^5)z^5
\nn \\
&+& 2 (2+ 21r + 189 r^2 + 964 r^3 + 447 r^4 - 127 r^5)z^6
\nn\\
&-& 4 (2+ 16  r+ 91 r^2 + 86 r^3 - r^4) z^7
\nn\\
&+&  (7+46 r+ 48 r^2 + 11 r^3)  z^8
\nn\\
&-& \left.\left.  (3+3 r+ r^2) z^9  \right]
\ln\left({z-2r+\sqrt{z^2-4r}\over z-2r-\sqrt{z^2-4r}} \right)
\right\}.
\label{dsig:dz:long:pol:v2}
\eqa

\subsection{Angular-energy distribution for unpolarized $\bm{J/\psi}$
\label{sec:1}%
}

The doubly differential angular-energy  distribution of unpolarized
$J/\psi$ can be parameterized in the following form:
\bqa
{d \sigma^{(i)} \over dz d\cos\theta}\big[ e^+e^-\to J/\psi+gg\big]
&=&  S^{(i)}(z) \big[ 1 + A^{(i)}(z)\, \cos^2\theta \big],
\label{double:differential:X:section}%
\eqa
where the superscript $i=0, 2$ represents the leading-order and
first-order contributions in relativistic expansion.

The corresponding functions at LO in $v$ read:
\bqa
S^{(0)}(z) &=& {3\,\check{\sigma}_0 \over 4} \, {1\over (2-z)^2 (z -2r)^3 (z^2-4r)}
\nn \\
&\times& \left\{ (z-2r)\sqrt{z^2-4r} \,\left[\,-4r(1+r)(3+12r+13r^2)
\right.\right.
\nn\\
&+& 32r(1+r)(1+3r)z +4(1-7r-12 r^2+2r^3) z^2
\nn \\
&-& \left. 4(2+r+3r^2)z^3 +7(1+r) z^4 - 2\,z^5\,\right]
\nn\\
&-& 2(1+r-z)\,\left[\,4r^2(1-r) (3+24r+13r^2)- 8r^2 (7-3r-12 r^2)z
\right.
\nn \\
&+& 2\,r(1-10 r-27 r^2+4 r^3) z^2+ 2r (7+7r-6 r^2) z^3
\nn \\
&-& \left. \left. (1-r)(1+5r) z^4 \right]\,
\ln\left({z-2r+\sqrt{z^2-4r}\over z-2r-\sqrt{z^2-4r}}\right)
\right\},
\label{dsig:dz:dcosth:S:v0}
\eqa
and
\bqa
S^{(0)}(z) A^{(0)}(z) &=&  {3\,\check{\sigma}_0 \over 4} \,{1 \over (2-z)^2 (z -2r)^3 (z^2-4r)}
\nn\\
&\times& \left\{ (z-2r)\sqrt{z^2-4r} \,\left[\,4r(1+5r+19r^2+7r^3)
\right.\right.
\nn\\
&-& 96 r^2(1+r)z -4(1-5r-22 r^2-2r^3) z^2
\nn \\
&-& \left. 4r(7+3r)z^3+(5+7r) z^4 - 2\,z^5 \right]
\nn \\
&+& 2(1+r-z)\,\left[\,4r^2(1+7r) (1-r^2)
 - 8r^2 (1+3r)(1-4r)z \right.
\nn \\
&-& 2r(1+10 r+57 r^2+4 r^3) z^2+ 2r (1+29r+6 r^2) z^3
\nn \\
&+& \left.\left. (1-8r-5r^2) z^4 \right]\,
\ln\left({z-2r+\sqrt{z^2-4r}\over z-2r-\sqrt{z^2-4r}}\right)
\right\}.
\label{dsig:dz:dcosth:SA:v0}
\eqa
Note these expressions are exactly twice smaller than Eqs.~(A1a) and
(A1b) in Ref.~\cite{Cho:1996cg}.

At the relative order $v^2$, the corresponding functions $S(z)$
and $A(z)$ are
\bqa
& & S^{(2)}(z)  = {\check{\sigma}_0
\, \langle v^2
\rangle_{J/\psi}\over 4}\, %
{1\over (2-z)^4 (z -2 r)^5(z^2-4r)}
\nn \\
&\times& \left\{ (z-2 r)\sqrt{z^2-4 r} \left[\,-64 r^2 (9+35r-8 r^2-14
r^3-57 r^4-45 r^5) \right.\right.
\nn \\
&+& 32 r^2 (64+ 84 r - 89 r^2 - 269 r^3 - 275 r^4 -  19 r^5) z
\nn \\
 &+ &
16 r (13  -69 r + 61 r^2 + 557 r^3 + 638 r^4 + 64 r^5) z^2
\nn \\
&-& 8 r (110+161 r + 559 r^2 + 725 r^3+ 3 r^4 - 14 r^5) z^3
\nn \\
&+& 4r (287 + 451 r + 383 r^2 - 251 r^3 -70 r^4) z^4
\nn \\
&+& 8 (2 - 87 r -29 r^2 + 109 r^3 + 35 r^4) z^5
\nn \\
&-& 4(6 -32 r+ 75 r^2 + 35 r^3)z^6
\nn \\
&+&  \left. 2 (9+ 9 r + 16 r^2) z^7 - (5-r) z^8 \right]
 \nn\\
 &+& 2 \left[\,-64
r^3 (9 + 38 r + 13 r^2 + 18 r^3 + 53 r^4 + 80 r^5 + 45 r^6) \right.
\nn \\
&+& 32 r^3 (70 + 166 r + 123 r^2 + 340 r^3 + 620 r^4 + 390 r^5+ 19
r^6) z
\nn \\
&+& 16 r^2 (19 - 140 r- 430 r^2- 922 r^3 - 1927 r^4 - 1410 r^5 - 102
r^6) z^2
\nn \\
&-& 8 r^2 (124 - 219 r - 1406 r^2 - 3154 r^3 - 2596 r^4 - 139 r^5 +
14 r^6) z^3
\nn \\
&-& 4 r (18- 193 r + 840 r^2 + 2976 r^3 + 2400 r^4 - 327 r^5 - 98
r^6) z^4
\nn\\
&+& 8r(29+ 28 r+ 355 r^2 +181 r^3 - 352 r^4 - 77 r^5)z^5
\nn \\
&+& 2 (2- 141 r - 131 r^2 + 293 r^3 + 1041 r^4 + 272 r^5)z^6
\nn \\
&-& 4 (2 - 34 r + 71 r^2 + 196 r^3 + 71 r^4) z^7
\nn\\
 &+& (7 - 5 r+ 147 r^2 + 87 r^3)  z^8
\nn \\
& - & \left.\left.  (3+ r + 14 r^2) z^9 \right]
\ln\left({z-2r+\sqrt{z^2-4r}\over z-2r-\sqrt{z^2-4r}} \right)
\right\},
\label{dsig:dz:dcosth:S:v2}
\eqa
and
\bqa
& & S^{(2)}(z) A^{(2)}(z) = {\check{\sigma}_0  \, \langle v^2
\rangle_{J/\psi}\over 4}\, %
{1\over (2-z)^4 (z -2 r)^5(z^2-4r)}
\nn \\
&\times& \left\{
 (z-2
r)\sqrt{z^2-4 r} \left[\,64 r^2 (3+ 17 r-8 r^2-10 r^3- 11 r^4- 15 r^5)
\right.\right.
\nn \\
&-& 32 r^2 (16+ 12 r - 99 r^2 - 79 r^3 - 113 r^4 -  r^5) z
\nn \\
 &- &
16 r (7  + 89 r + 271 r^2 + 335 r^3 + 370 r^4 + 32 r^5) z^2
\nn \\
& + & 8 r (90+ 595 r + 789 r^2 + 775 r^3+ 161 r^4 + 14 r^5) z^3
\nn \\
& - &  4 (1+r) (8 + 325 r + 836 r^2 + 321 r^3 + 70 r^4) z^4
\nn\\
&+& 8 (10 + 129 r + 291 r^2 + 141 r^3 + 35 r^4) z^5
\nn \\
&-& 4 (18 + 104 r+ 119 r^2 + 35 r^3)z^6
\nn\\
&+& \left. 2 (1+r)(17  + 16 r) z^7 - (7-r) z^8 \right]
\nn\\
&+& 2 \left[\, 64 r^3 (3 + 18 r + 15 r^2 + 30 r^3 - 25 r^4 + 8 r^5 +
15 r^6) \right.
\nn \\
&-& 32 r^3 (18 + 98 r + 201 r^2 + 28 r^3 + 36 r^4 + 162 r^5+ r^6) z
\nn \\
& - & 16 r^2 (9 - 12 r- 490 r^2- 566 r^3 - 389 r^4 - 710 r^5 - 34
r^6) z^2
\nn \\
&+& 8 r^2 (36 - 257 r - 1410 r^2 - 1670 r^3 - 1980 r^4 - 193 r^5 -
14 r^6) z^3
\nn \\
&+& 4 r (14 + 221 r + 1360 r^2 + 3208 r^3 + 3800 r^4 +  595 r^5 + 98
r^6) z^4
\nn\\
& - & 8r (35+ 300 r+ 817 r^2 + 1203 r^3 + 284 r^4 + 77 r^5)z^5
\nn \\
&+& 2 (2 + 219 r + 1177 r^2 + 1973 r^3 + 669 r^4 + 272 r^5)z^6
\nn\\
&-& 4 (2 + 82 r + 275 r^2 + 124 r^3 + 71 r^4) z^7
\nn\\
& + & (7 + 119 r + 119 r^2 + 87 r^3)  z^8
\nn \\
& - &  \left.\left. (3+ 5 r + 14 r^2) z^9 \right]
\ln\left({z-2r+\sqrt{z^2-4r}\over z-2r-\sqrt{z^2-4r}} \right)
\right\}.
\label{dsig:dz:dcosth:SA:v2}
\eqa

Integrating Eq.~(\ref{double:differential:X:section}) over the polar angle $\theta$
from $0$ to $\pi$, we arrive at the following identity:
\bqa
{d \sigma^{(i)}\over dz} \big[e^+e^-\to J/\psi+gg \big]= 2
S^{(i)}(z)\left[1+{1\over 3} A^{(i)}(z)\right],
\eqa
where $d \sigma^{(i)}/dz$ represent the energy distributions
for unpolarized $J/\psi$, which have been given in
Eqs.~(\ref{dsig:dz:unpol:v0}) and (\ref{dsig:dz:unpol:v2}). This
relation can serve as a consistency check of our results.
We have explicitly verified that our expressions obey this relation
for both $i=0$ and $2$.

\section{Equivalence between our matching method and
the ``orthodox" one
\label{equiv:mine:matching:vs:orthdox}%
}

It is curious to ask, whether the inclusive $J/\psi$ production rate
derived from our matching procedure, can be translated into a more
orthodox form, that is, everything is expressed in terms of charm
quark mass rather than the charmonium mass. That corresponds to what
would be obtained from a literal matching method. As we shall see,
this is possible only for the {\it integrated} cross section. And we
like to stress, there should be no any theoretical ambiguity and confusion
for the relativistic correction contribution at this level.

To better orientate ourselves, let us begin with a one-dimensional toy
integral:
\bqa
\int^{f(x,y)}_{g(x,y)} dt \, W(t,y) &=& \int^{f(x,0)}_{g(x,0)} dt \,
W(t,0) + y \left\{ \int^{f(x,0)}_{g(x,0)} dt \, W'_y(t,0) \right.
\nn \\
&+&  \left.  W\big(f(x,0),0 \big) \, f'_y(x,0)
-W\big(g(x,0),0\big)\, g'_y(x,0) \right\} + O(y^2),
\label{taylor:expansion:identity}
\eqa
where we have assumed the integrand $W$ is regular at the end points of
the integral and used the shorthand $f'_y(x,0)\equiv
\partial f(x,y)/\partial y|_{y=0}$.
The final result of the integral in the left-hand side will be a
function of $x$ and $y$. Here $y$ is assumed to be a small constant,
and it is assumed that both the integrand $W$ and the integration
boundaries $f$, $g$ depend on $y$. Our goal is to reexpress the
original integral in a Taylor-series in $y$. Since in many
situations, the closed form for such integral is presumably not
available, or at least difficult to obtain,
it is thus desirable to find a general numerical recipe to
accomplish this expansion.

In the right-hand side of (\ref{taylor:expansion:identity}), we give
the intended answer for this expansion through the linear order in
$y$. The leading term is obtained by neglecting $y$ simultaneously
in the integrand and integration boundaries. The coefficients of
order $y$ come from either expanding the integrand or taking into
account the correction to the integration boundaries.

The goal is to reexpress our ``leading-order" cross
sections in terms of a new series including the first-order relativistic
correction, with all the occurrences of $M_{J/\psi}$ replaced by
$2m_c$ in a consistent way. Note both the matrix element squared
and the boundaries of the phase space integral depend on
$v^2$ implicitly through $M_{J/\psi}$. Clearly, $v^2$ is the counterpart of $y$ in
(\ref{taylor:expansion:identity}) that acts
as the small expansion parameter. One can utilize
(\ref{taylor:expansion:identity}) to work out the desired expanded form.

For concreteness, we take the unpolarized $J/\psi$ energy
distribution as an example. The LO energy spectrum of $J/\psi$
derived from our matching method has been given in
Eq.~(\ref{dsig:dz:unpol:v0}):
\bqa
&& {d \sigma^{(0)}\over dz} \big[e^+e^-\to J/\psi+gg \big]
\nn\\
& = & {256 \pi (e_c\alpha\alpha_s)^2  \over 27 M_{J/\psi}\,s^2}
\,\langle {\mathcal O}_1^{J/\psi} \rangle\, {1\over (2-z)^2 (z
-2r)^3}
\nn \\
&\times & \left\{ (z-2r)\sqrt{z^2-4 r}\,
\bigg[ 4(1+5 r+7r^2+4 r^3) \right.
\nn \\
&-& 12(1+r)(1+2r) z +(13+14r)z^2 -4z^3\bigg]
\nn\\
&+& 4(1+r-z)\,\bigg[ 2r(1-r)(1+8r+4 r^2)- 2r(5-2r-6 r^2) z
\nn \\
&+& \left.(1+r-5r^2)z^2 \bigg]
\ln\left({z-2r+\sqrt{z^2-4r}\over z-2r-\sqrt{z^2-4r}}\right)
\right\}.
\label{upolar:jpsi:distrib}%
\eqa
For clarity, here we abandon the use of the abbreviation $\check{\sigma}_0$ and
supply the complete expression of the prefactor.

In accordance with (\ref{taylor:expansion:identity}), we may
reexpress the integrated cross section of
(\ref{upolar:jpsi:distrib}) as a sum of the following three terms,
each of which now depends on the charm quark mass rather than the
$J/\psi$ mass:
\bqa
\int_{2\sqrt{r}}^{1+r} dz {d \sigma^{(0)} \over dz} &=&
\int_{2\sqrt{r_0}}^{1+r_0} {d \tilde{\sigma}^{(0)} \over dz}+
\int_{2\sqrt{r_0}}^{1+r_0} {d \tilde{\sigma}^{(2a)}\over dz}+\tilde{
\sigma}^{(2b)}+O(v^4\sigma).
\label{equiv:mine:vs:orthodox}
\eqa
where $r_0 \equiv {4m_c^2\over s}$. Upon expanding
(\ref{upolar:jpsi:distrib}), we need replacing every occurrence of
$M_{J/\psi}$ with the combination of $m_c$ and $\langle v^2
\rangle_{J/\psi}$ through the G-K relation (\ref{G-K:rel}):
\begin{subequations}
\bqa
r &=& r_0 \big[ 1+ \langle v^2 \rangle_{J/\psi}+O(v^4) \big],
\\
{1\over M_{J/\psi}} &=& {1\over 2m_c}\left(1-{1\over 2}\langle v^2 \rangle_{J/\psi}+O(v^4) \right).
\eqa
\end{subequations}
In the resulting new expression,
we only need retain those terms at most of order $v^2$.

The first term in the right side of (\ref{equiv:mine:vs:orthodox}) constitutes the
leading contribution, the second one comes from the expansion of the
integrand, and the third one arises from the correction due to
integration boundaries. Their explicit expressions are
\begin{subequations}
\bqa
{d \tilde{\sigma}^{(0)}  \over dz} & = &  \left.  {d \sigma^{(0)}
\over dz} \right|_{M_{J/\psi}\to 2m_c,\; r\to r_0},
\\
{d \tilde{\sigma}^{(2a)} \over dz}  &=&
 {64 \pi
(e_c\alpha\alpha_s)^2  \over 27   m_c s^2} \langle {\cal
P}_1^{J/\psi}\rangle {1\over (2-z)^2 (z -2r)^4\sqrt{z^2-4r}}
\label{sigma:2a}
\\
&\times & \left\{ (z-2 r)\left[\,-32 r^2 (5+17 r - 4 r^2 - 12
r^3)\right.\right.
\nn\\
&-& 16 r (1-17 r - 8 r^2 + 32 r^3)z
\nn \\
&+&  8 r (15 + 7 r + 17 r^2 - 12 r^3)z^2
\nn\\
&-&  4 (3 + 47 r + 11 r^2 - 32 r^3)z^3
\nn \\
&+& \left. 2 (10 + 51 r - 14 r^2) z^4 - 13(1+2r)z^5 + 4z^6\right]
\nn\\
&+& 4\sqrt{z^2-4r} \left[\,4r^2(5+24 r+ 3 r^2 + 8 r^3 + 12 r^4)
\right.
\nn \\
&+& 2r (1-36 r- 45 r^2 - 32 r^3 - 56 r^4) z
\nn \\
&+& 2r (1+ 30 r + 36 r^2 + 53 r^3) z^2
\nn\\
&-&(1+ 2r + 34 r^2 + 55 r^3) z^3
\nn \\
&+& \left.\left.\left.  (1-  r + 15 r^2) z^4 \right]
\ln\left({z-2r+\sqrt{z^2-4r}\over z-2r-\sqrt{z^2-4r}}\right)
\right\}\right|_{r\to r_0},
\nn\\
\tilde{\sigma}^{(2b)} &=& \left.
 {512 \pi
(e_c\alpha\alpha_s)^2  m_c  \over 27 s^3} \langle {\cal
P}_1^{J/\psi}\rangle  {1+2r\over 1-r}\right|_{r\to r_0},
\label{sigma:2b}
\eqa
\end{subequations}
where $\langle {\cal
P}_1^{J/\psi}\rangle$ is given in (\ref{P1:jpsi:vac:ME}).
Needless to say, the new LO term is exactly of the same
functional form as the old one in (\ref{upolar:jpsi:distrib}),
except $M_{J/\psi}$ everywhere replaced by $2 m_c$.
For the newly generated relativistic correction pieces
$\tilde{\sigma}^{(2a)}$ and $\tilde{\sigma}^{(2b)}$, one
does not need to carefully distinguish $r_0$ and $r$ in them,
since the induced error would be of order $v^4$,
which is beyond the intended accuracy of this work.

All these three terms, in combination with (\ref{dsig:dz:unpol:v2}),
the genuine $O(v^2)$ contribution in our matching approach~\footnote{Note we can carelessly
replace $M_{J/\psi}$ by $2m_c$
in (\ref{dsig:dz:unpol:v2}), and, in the corresponding phase-space integral boundaries,
since the induced error would be $O(v^4)$.},
constitute an alternative but equally valid prediction to the {\it
integrated} $J/\psi$ cross section that is accurate at relative
order $v^2$.
Since the expression for the integrated $J/\psi$ production rate, when everything
is expressed in term of $m_c$, has no any ambiguity through $O(v^2)$, it can be
used to check the correctness of the calculation performed in an
``orthodox" matching method ({\it e.g.}, see~\cite{He:2009uf}).

To clearly see how (\ref{equiv:mine:vs:orthodox}) works, we can take advantage of our
analytic knowledge for the integrated $J/\psi$ cross section at $O(v^0)$.
Directly Taylor expanding (\ref{sig:int:unpol:v0}) around $r=r_0$ to first order in $r-r_0$,
we find
\bqa
\label{sig:int:unpol:v2:corr}
 & & \tilde{\sigma}^{(2a)}+ \tilde{
\sigma}^{(2b)}
=  -{128 \pi (e_c\alpha\alpha_s)^2 \over 27 m_c s^2}
\langle {\cal P}_1^{J/\psi}\rangle
\left\{{2-9 r+39 r^2 -28 r^3+8r^4 \over 4(1-r)^3} \right.
\nn \\
&\times & {\rm arctanh}^2\sqrt{1-r}+ {3-4r-9 r^2+4r^3 \over (1-r)^{5/2}}\,{\rm arctanh}\sqrt{1-r}
\nn\\
&+ & \left.  {5-11 r+ 33r^2 -3r^3\over 4 (1-r)^3} \ln r -{11- 19 r - 46 r^2 + 18 r^3 \over 4 (1-r)^2}\right\}.
\eqa
We have numerically compared (\ref{sig:int:unpol:v2:corr}) with
the sum of (\ref{sigma:2a}) and (\ref{sigma:2b}) upon integration over
the full range of $z$, and indeed found exact agreement.

We have also numerically checked that, both sides of (\ref{equiv:mine:vs:orthodox}), assuringly,
do agree with each other at the integrated level,
up to an error of order $v^4$~\footnote{For $J/\psi$ production at
$B$ factory, the relativistic correction stemming from expanding the
phase space boundaries, $\tilde{\sigma}^{(2b)}$, makes negligible contribution
due to the additional suppression by $m_c^2/s$.}.

Finally, it might be worth mentioning that, the differential distribution $d
\tilde{\sigma}^{(2a)}/dz$ diverges at both upper and
lower ends of $z$ (albeit being the integrable singularities):
\begin{subequations}
\bqa
\left. {d \tilde{\sigma}^{(2a)} \over dz}\right|_{z\to 2 \sqrt{r}}
&\longrightarrow & -{64 \pi (e_c\alpha\alpha_s)^2 \over 27 m_c s^2}
\langle {\cal P}_1^{J/\psi}\rangle {4-8\sqrt{r}+7r \over
(1-\sqrt{r})^2 \sqrt{z^2-4r}},
\\
\left. {d \tilde{\sigma}^{(2a)} \over dz} \right|_{z\to 1+r}
&\longrightarrow& -{64 \pi (e_c\alpha\alpha_s)^2 \over 27 m_c s^2}
\langle {\cal P}_1^{J/\psi} \rangle
\nn\\
&\times& {(1-3r)(1 + 11 r +6 r^2)+8 r (1 - 3 r - r^2)
\ln\big[{r\,(1+r-z)\over (1-r)^2}\big] \over (1-r)^3}.
\eqa
\end{subequations}
These artificial end-point singularities affiliated with the
relativistic correction to $J/\psi$ energy distributions,
especially the one appearing at the lower end,
are clearly at odds with one's expectation
and certainly not favored by the data.
This may signal that, even if feasible,
it is not of much benefit to perform the NRQCD matching
in a strictly orthodox ansatz.
Instead the matching method described in this work
seems much more satisfactory.


\end{document}